\documentclass[nonatbib]{elsarticle}
\usepackage[utf8]{inputenc}
\usepackage[british]{babel}
\usepackage{csquotes}
\usepackage{subfiles}
\usepackage{booktabs}
\usepackage{graphicx}
\usepackage{multirow}
\usepackage{lscape}
\usepackage{setspace}
\usepackage{lscape}
\usepackage{eurosym}
\usepackage[labelfont=bf, font=footnotesize]{caption}
\DeclareCaptionLabelSeparator{vline}{$~\vert~$}
\usepackage[table,xcdraw]{xcolor}
\usepackage[hidelinks]{hyperref}
\hypersetup{colorlinks=true,citecolor={blue},citebordercolor={blue}}
\usepackage{geometry}
\makeatletter
\let\c@author\relax
\makeatother
\usepackage[backend=biber, style=numeric-comp, autocite=superscript, maxcitenames=1, isbn=false, date=year, url=false, sorting=none]{biblatex}
\DeclareAutoCiteCommand{superscript}{\supercite}{\supercites}
\AtEveryBibitem{\clearfield{note} \clearlist{language}}
\makeatletter
\def\ps@pprintTitle{%
  \let\@oddhead\@empty
  \let\@evenhead\@empty
  \let\@oddfoot\@empty
  \let\@evenfoot\@oddfoot}
\makeatother

\title{A moderate share of V2G outperforms large-scale smart charging of electric vehicles and benefits other consumers\tnoteref{t1}}

\author[1,2,3]{Adeline Guéret\corref{cor1}}

\author[1]{Carlos Gaete-Morales}

\author[1]{Wolf-Peter Schill}

\cortext[cor1]{Corresponding author: \texttt{agueret@diw.de}.}
\affiliation[1]{organization={DIW Berlin},
addressline={Anton-Wilhelm-Amo-Straße 58},
city={10117 Berlin},
country={Germany}}

\affiliation[2]{organization={Technische Universität Berlin},
addressline={Straße des 17.\ Juni 135},
city={10623 Berlin},
country={Germany}}

\affiliation[3]{organization={OFCE - Sciences Po Paris},
addressline={10, Place de Catalogne},
city={75014 Paris},
country={France}}

\begin{document}


\begin{abstract}
While battery electric vehicles (BEVs) play a key role for decarbonizing the transport sector, their impact on the power sector heavily depends on their charging strategies. Here we systematically analyze various combinations between inflexible, smart and bidirectional (or vehicle-to-grid, V2G) charging of 15 million electric cars in Germany. Using a capacity expansion model, we find that even a moderate share of bidirectional charging below 30\% leads to lower system costs than a fully smartly charging BEV fleet. At a V2G share of 50\%, costs are even lower than in a system without any BEVs. This means that the flexibility effect of half of the BEV fleet charging bidirectionally outweighs the demand effect of the whole BEV fleet. We show how costs savings are driven by the ability of V2G to serve demand, especially during hours with high residual load. We also explore the distributional effects of respective electricity price changes. While V2G car owners internalize a substantial share of overall cost savings, the benefits increasingly spill over to other electricity consumers as the share of bidirectional charging grows. We conclude that policymakers should focus on enabling a moderate fleet share of V2G rather than on enabling every car to charge smartly.
\end{abstract}

\doublespacing

\maketitle
\renewcommand{\thefootnote}{\alph{footnote}}

\section{Introduction}

Battery electric vehicles (BEVs) are an effective way to reduce greenhouse gas emissions in the transportation sector, especially in passenger road transportation\autocite{van2025demand, van2025electricity, speizer2024integrated, rottoli2021alternative, knobloch2020net}, in combination with increasingly decarbonizing power sectors\autocite{simaitis2025battery, zhang2020role, hoekstra2019underestimated}. The switch from internal combustion engine vehicles (ICEVs) to BEVs inevitably impacts power systems, as it means that the electricity supply must meet a larger overall demand. According to current projections, the number of light-duty BEVs on the roads in Europe could reach 35~million by 2030, compared to 8.7~million in 2024\autocite{global_ev_data_explorer_2025}. In Germany alone, there is a 15~million BEVs target for 2030 compared to a stock of 1.9~million in August 2025\autocite{roth_schill_oet_2022}. Under this fleet size target, BEVs could increase today's German electricity consumption by up to 10\% by 2030.

Previous research has shown that the overall impact of coupling light-duty vehicles to the power sector strongly depends on the charging strategy considered\autocite{frank2025potential, bogdanov2024role, lauvergne2022integration, powell2022charging, taljegard2019impacts, hanemann2017effects, schill2015power, richardson2013electric}. Inflexible charging can lead to a significant peak load increase\autocite{kamana2024driving, fischer2019electric, muratori2018impact}. Smart charging, also referred to as controlled, coordinated, optimized, or unilateral charging, allows to better align the timing of BEV charging with the availability of low-cost electricity. Hence, smart charging can limit the increase in residual load peaks triggered by the introduction of BEVs to a significant extent\autocite{crozier2020opportunity}, reducing system costs increases\autocite{mangipinto2022impact, szinai2020reduced}. Besides, smartly charging electric vehicles could help integrate variable renewable energy sources (vRES) if charging is aligned with episodes of negative residual load\autocite{dallinger2012grid, juul2011optimal, ekman2011synergy}. This can reduce curtailment and/or increase the optimal installed variable renewable capacity\autocite{heuberger2020ev, szinai2020reduced, loisel2014large}. 
Bidirectional charging or vehicle-to-grid (V2G) could reduce overall system costs further and increase the power sector's stability and reliability, even compared to a power sector without BEVs. While early contributions have established this general point\autocite{kempton2005vehicle2, kempton2005vehicle, kempton1997electric}, more recent articles have investigated this question with more detailed methods\autocite{wei2022planning, yao2022economic, wu2021benefits, taljegard2019impact, hanemann2018effects, druitt2012simulation, lund2008integration}. These positive effects are leveraged by using BEVs as distributed grid storages that can absorb excess cheap energy and feed it back to the grid in hours with high residual load, reducing the stationary storage capacity needed in the system\autocite{syla2025assessing, xu2023electric, brown2018synergies, forrest2016charging, tarroja2016assessing}. 

Here, we systematically investigate the impact of different charging strategies on the power sector, considering any possible combination between inflexible, smart and bidirectional charging in the vehicle fleet at 10\% increments. We quantify changes in system costs, optimal capacity and dispatch outcomes, and shed light on the distributional effects of various degrees of BEV flexibility on different electricity consumer groups. For this, we use an open-source power sector model of central Europe\autocite{zerrahn2017long}, calibrated for a 2030 case study, and focus on the rollout of BEVs in Germany. BEV time series are generated with an open-source tool\autocite{gaete-morales_open_2021} using empirical mobility data from a representative national travel survey.

To the best of our knowledge, such a systematic analysis of mixed charging strategies in the BEV fleet has not been undertaken so far, except in highly stylized frameworks that do not model policy-relevant electric mobility scenarios and power sectors\autocite{dioha2022idealized}. Previous analyses typically rely on extreme case scenarios, where most or all vehicles charge with the same strategy\autocite{strobel2022joint, wei2022planning, hanemann2017effects, loisel2014large}. Varying shares of a given charging strategy are sometimes investigated, but only for selected cases and considering two charging strategies at most\autocite{frank2025potential, bogdanov2024role, taljegard2019impact, brown2018synergies}. In contrast, our research design allows to investigate the interactions between all three charging strategies. We also analyze how the switch to more flexible BEV operations impacts different electricity consumer groups, which has not been done for mixed charging strategies before\autocite{emelianova2025welfare}. 

We find that a moderate share below 30\% of BEVs charging bidirectionally can already keep the overall system costs increase at levels similar to a fully smartly charging BEV fleet. If half of the fleet charges bidirectionally, total costs are even lower than in a power sector without BEVs. Owners of bidirectionally charging BEVs benefit substantially from the services they provide to the grid. However, we find that with increasing shares of V2G these benefits increasingly spill over to non-BEV consumers, whose electricity bills decrease.

\section{Results}
\label{sec:results}

\begin{figure}[!ht]
    \centering
    \includegraphics[width=0.5\linewidth]{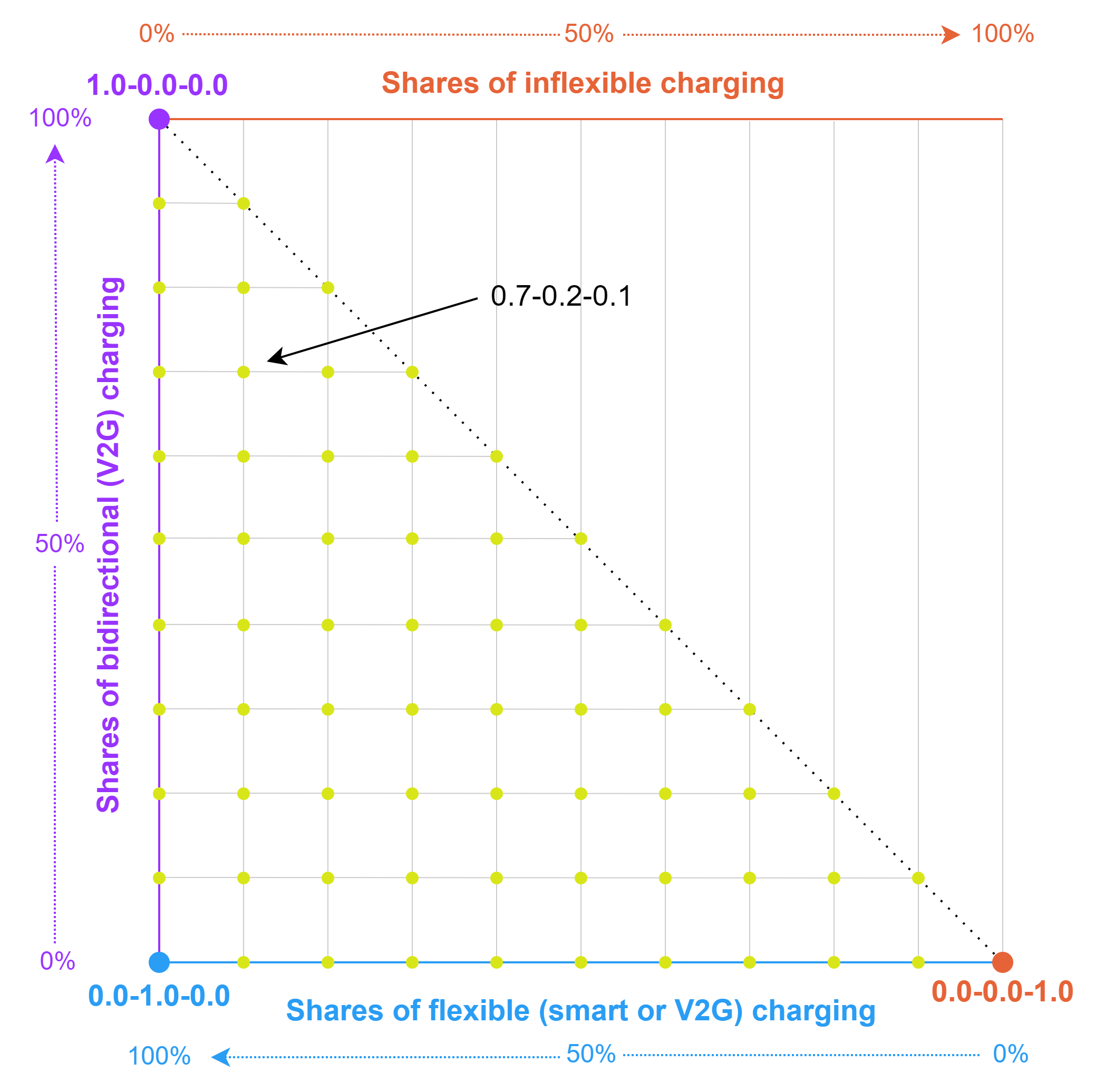}
    \caption{
    \textbf{Scenarios analyzed for a 15 million BEV fleet.} 
    Each dot corresponds to a scenario, i.e., a unique distribution of the BEV fleet across the three considered charging strategies. Bigger dots depict the three corner scenarios in which all BEVs have the same charging strategy. Together, the scenarios cover all possible combinations of charging strategies at 10\% increments.
    }
    \label{fig:scenarios}
\end{figure}

In what follows, we compare scenarios with 15 million BEVs in Germany to a reference without BEVs. In each scenario, we model a specific combination of inflexible charging, smart charging and bidirectional charging. Taken together, scenarios map all possible combinations of these three charging strategies at 10\% increments, which results in 66 scenarios (see Figure~\ref{fig:scenarios}). We label these scenarios with a series of three numbers whose order is: (1) the share of vehicles charging bidirectionally - (2) the share of smart charging vehicles - (3) the share of inflexible vehicles, e.g. ``$0.7-0.2-0.1$'' refers to 70\% charging bidirectionally, 20\% charging smartly, 10\% charging inflexibly. 

\subsection{Moderate shares of bidirectional charging substantially lower overall system costs}

\begin{figure}[!ht]
    \centering
    \includegraphics[width=.45\linewidth]{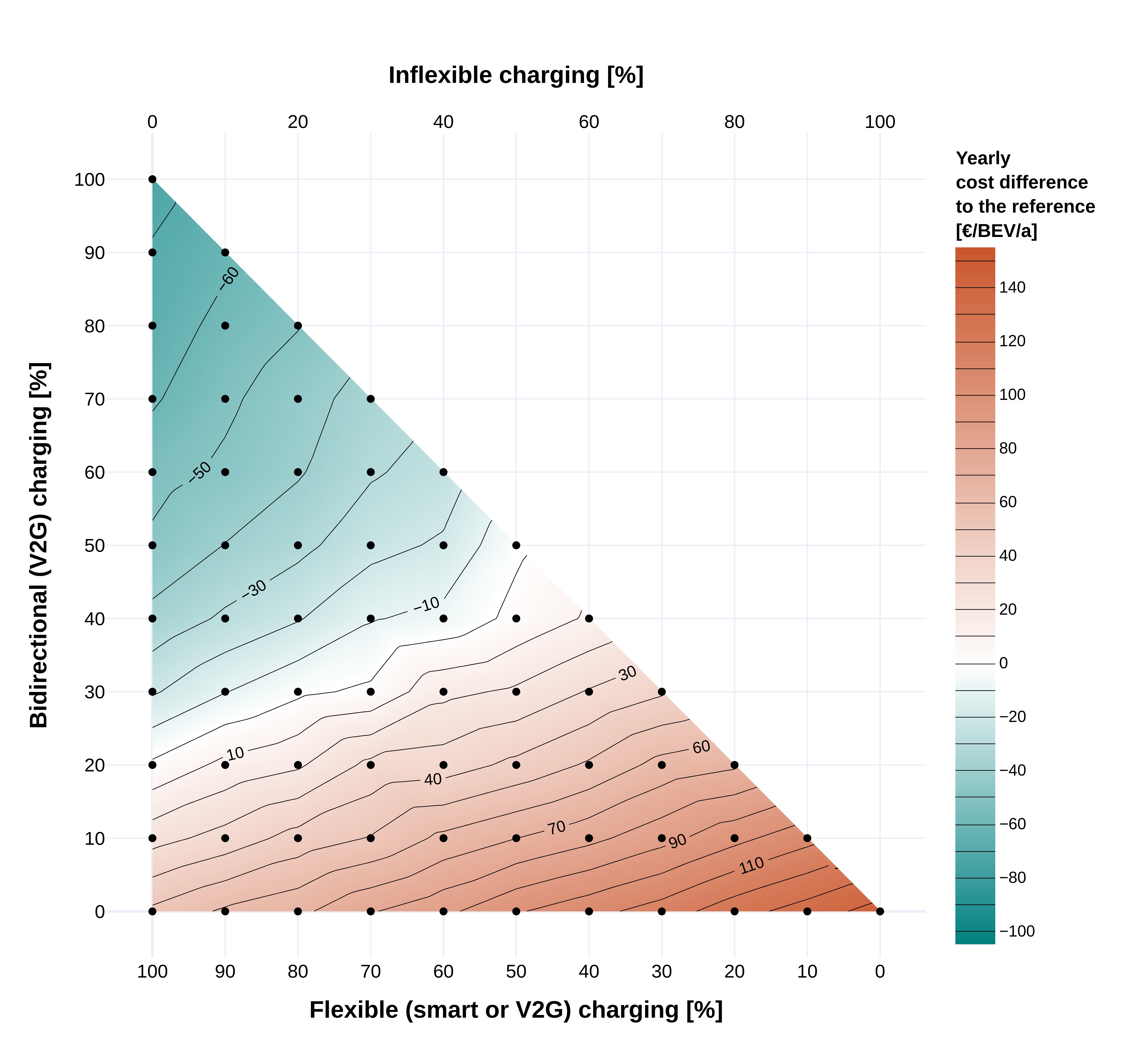}
    \includegraphics[width=.45\linewidth]{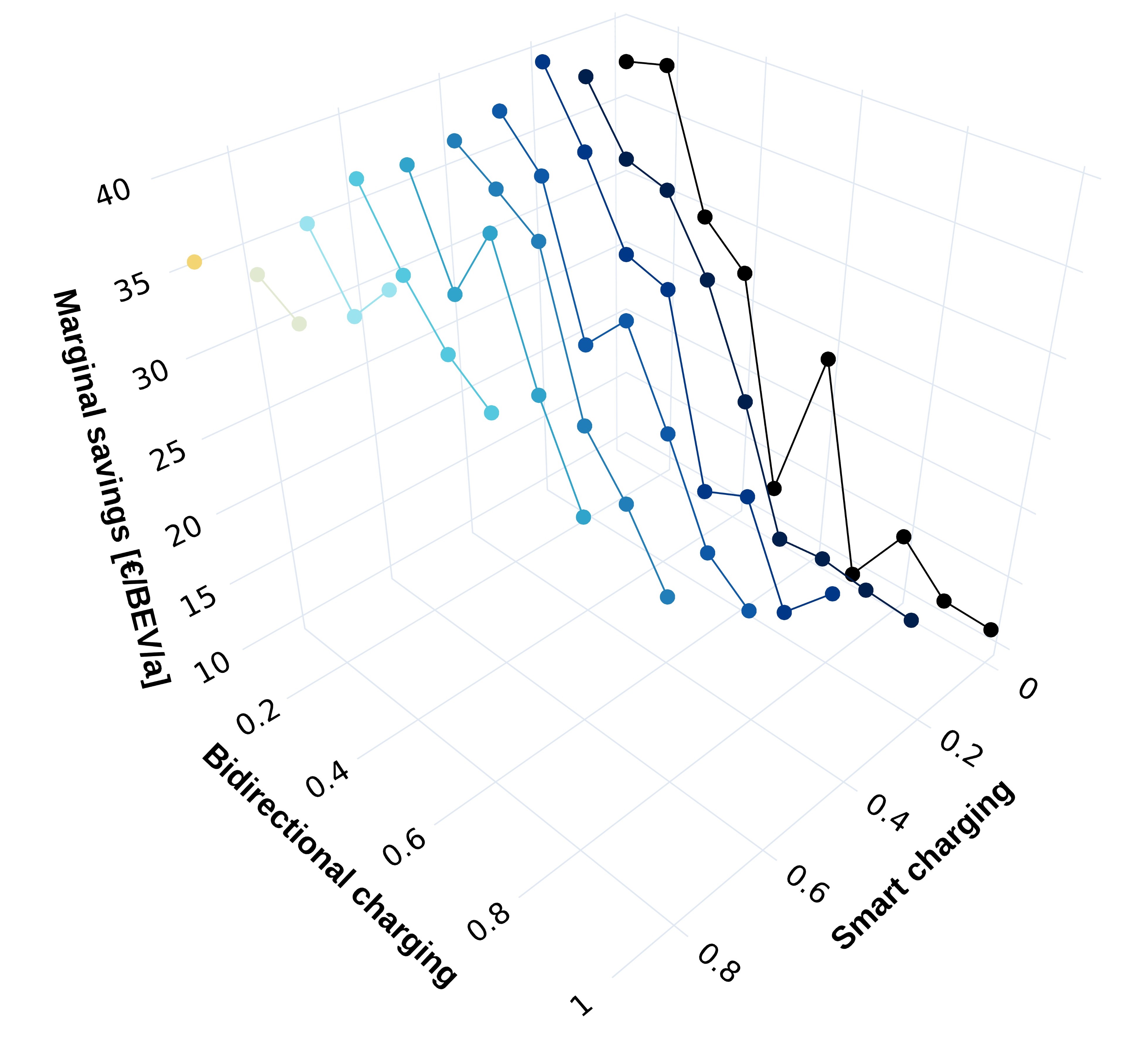}
    \caption{
    \textit{Left panel --} \textbf{Yearly system costs difference to the reference.} The reference refers to a setting without BEVs. Each dots represents a specific combination of charging strategies in the BEV fleet. Costs differences are provided per BEV per year, i.e., equally attributed to all BEVs in the fleet. 
    \textit{Right panel --} \textbf{Average marginal V2G share savings.} Color lines correspond to smart charging shares, from no smart charging (black) to full smart charging (yellow). Each dot represents system costs savings induced by increasing the share of bidirectionally charging BEVs by 10\% in the fleet, keeping the share of smartly charging BEVs constant. Savings are provided per BEV per year.}
    \label{fig:cost_difference}
\end{figure}

Our findings show that bidirectional charging entails much larger system costs savings than smart charging (Figure~\ref{fig:cost_difference}). Already a share of 30\% of bidirectionally charging cars in the fleet, combined with 70\% inflexibly charging cars, leads to lower system costs than when the entire BEV fleet charges smartly. 

We further find that when half of the BEV fleet engages in V2G, the overall system costs are even lower than in the case where there are no BEVs at all. With~50\% of BEVs charging bidirectionally and the remaining 50\% charging inflexibly (our setting labeled as ``0.5-0.0-0.5"), the cost difference to the reference is negative at around~2~euros per BEV per year. This means that the bidirectionally charging cars provide enough flexibility to the system to compensate the cost effect of the additional electricity demand of the BEV fleet. In other words, the \textit{flexibility effect} of bidirectionally charging~50\% of the BEV fleet outweighs the \textit{demand effect} of the whole BEV fleet. 

In contrast, if~50\% of the vehicles charge only smartly without V2G and the remaining~50\% charge inflexibly (``0.0-0.5-0.5''), costs increase by~99 euros per BEV per year compared to the reference without BEVs. Even with a fully smartly charging BEV fleet (``0.0-1.0-0.0''), the overall system costs increase by~52 euros per BEV per year, i.e., the \textit{demand effect} still outweighs the \textit{flexibility effect}. Hence, smart charging, even when rolled out in the entire BEV fleet, has a much lower potential for system costs benefits than bidirectional charging.  

The analysis further shows that it is more beneficial to convert smartly charging BEVs to vehicle-to-grid rather than enabling inflexibly charging BEVs to charge smartly. For most points in the lower half of Figure~\ref{fig:cost_difference} (left panel), moving upwards by~10\% (i.e., converting 10\% of smartly charging BEVs to V2G) brings larger cost savings than moving to the left by~10\% (i.e., making 10\% of inflexible BEVs charge smartly). For settings where all flexibly charging BEVs are already able to charge bidirectionally (outward diagonal of the triangle), increasing the number of bidirectionally charging BEVs (moving up and to the left) is also more beneficial than increasing the number of smartly charging BEVs (moving to the left).

However, the marginal benefits of V2G are decreasing. The highest average marginal cost savings of increasing the share of bidirectional charging by~10\% in the overall BEV fleet occur at low V2G shares, irrespective of how many BEVs are charging smartly. Figure~\ref{fig:cost_difference} (right panel) shows average marginal savings of increasing the share of bidirectional charging by~10\% in the fleet. Each line corresponds to a constant share of smartly charging BEVs. For any share of smartly charging BEV, increasing the share of V2G from zero to~10\% yields marginal savings of around~34-39 euros per BEV per year, even when already~90\% of the BEV fleet charges smartly. Average marginal savings then steeply decrease with the overall V2G share, but always remain positive.  

\subsection{Vehicle-to-grid integrates low-cost variable renewables more efficiently}

The introduction of electric cars with their respective charging strategies changes both optimal capacity and dispatch outcomes, which drive the costs effects discussed above.

\begin{figure}[!ht]
    \centering
    \includegraphics[width=.45\linewidth]{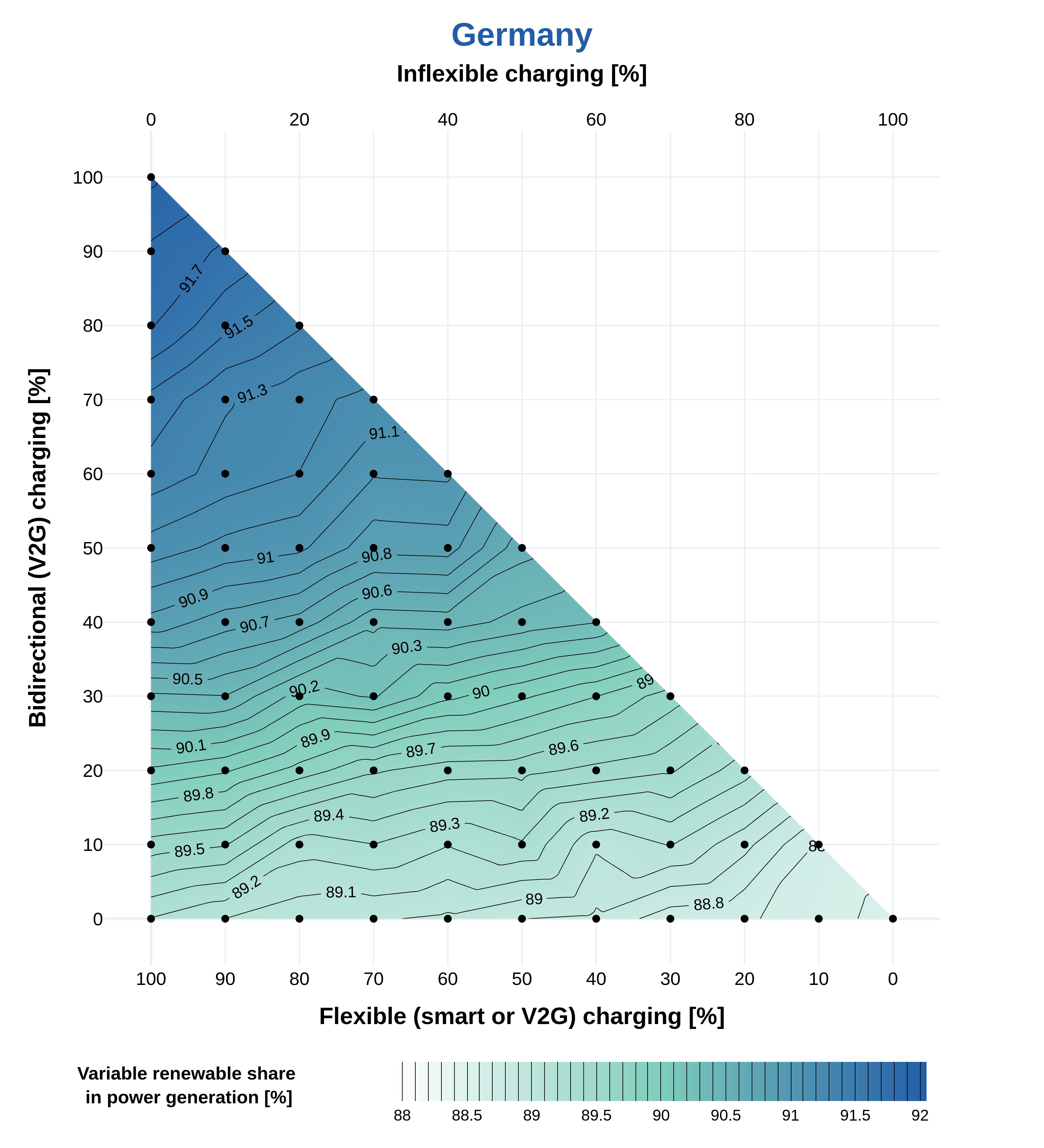}
    \includegraphics[width=.45\linewidth]{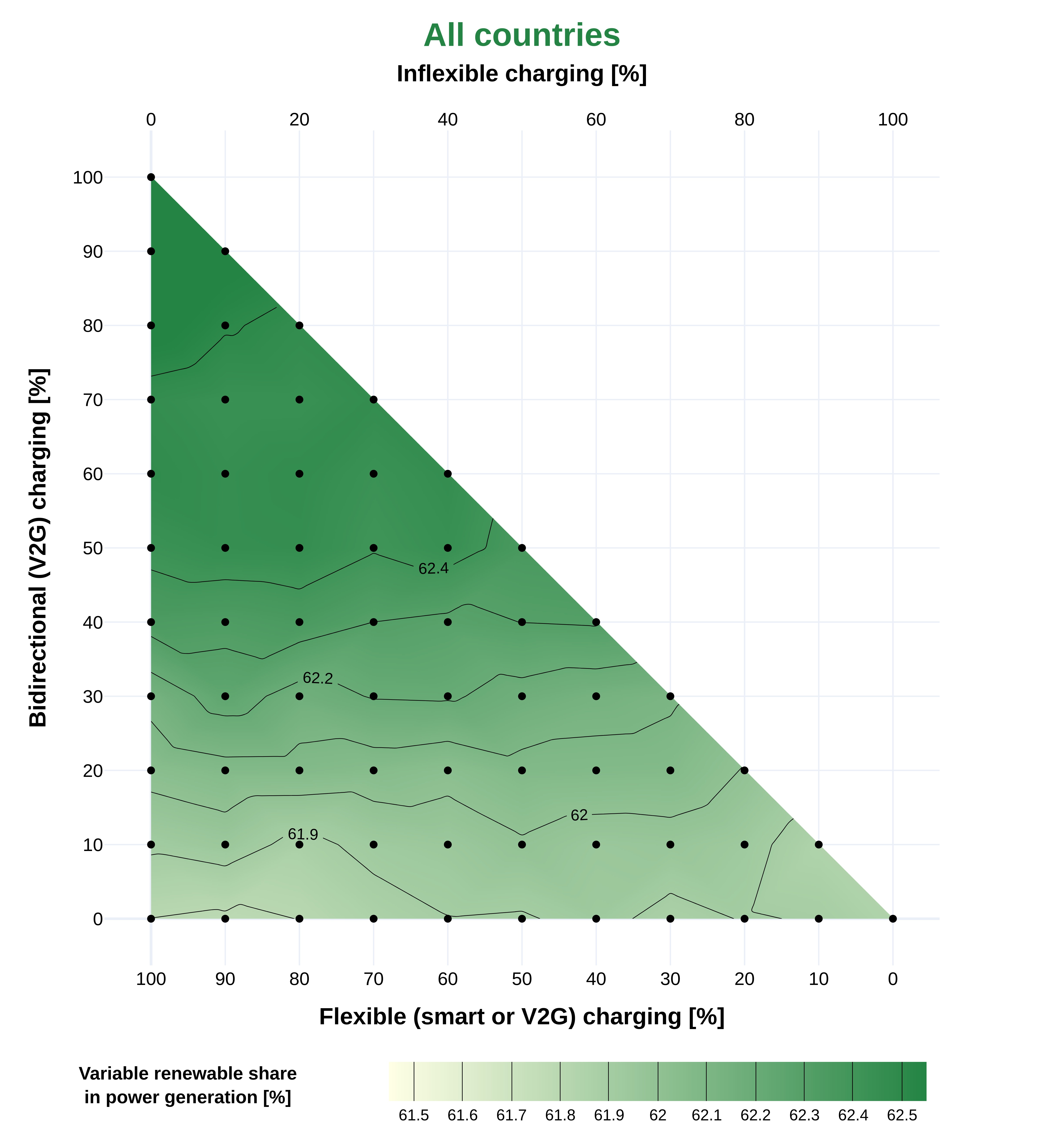}
    \caption{
    \textit{Left panel --} \textbf{Share of electricity generated from variable renewable energy sources (vRES) in Germany's total power generation.} 
    \textit{Right panel --} \textbf{Share of electricity generated from vRES in the total power generation of all countries.} vRES refer to run-of-river hydro, solar photovoltaic, wind onshore and wind offshore and vRES electricity generation refers to generation net of curtailed vRES generation.}
    \label{fig:res-shares}
\end{figure}

Note that in our model parametrization, vRES technologies have a lower levelized cost of electricity (LCOE) than other dispatchable technologies such as natural gas power plants (OGCT and CCGT). However, larger amounts of vRES are increasingly difficult to integrate in the power sector due to a growing temporal mismatch between additional generation and the residual load. This means that vRES technologies increasingly require flexibility technologies, such as short- and long-duration storage technologies, that help balance their variability. Cost-minimizing models hence typically find an optimal share of vRES, which depends on the relative costs of different generation and flexibility technologies.       

In such a setting, the introduction of flexibly charging BEVs increases the flexibility potential of the power sector, which leads to a higher optimal share of vRES. As depicted in Figure~\ref{fig:res-shares} (left panel), the variable renewable share in the aggregated power generation of Germany increases markedly when more vehicles charge bidirectionally. The graph also shows that increasing the V2G share drives up the variable renewable share much more than increasing the share of smart charging. When considering all countries, this effect is smaller (Figure~\ref{fig:res-shares}, right panel). 

\begin{figure}[!ht]
    \centering
    \includegraphics[width=1\linewidth]{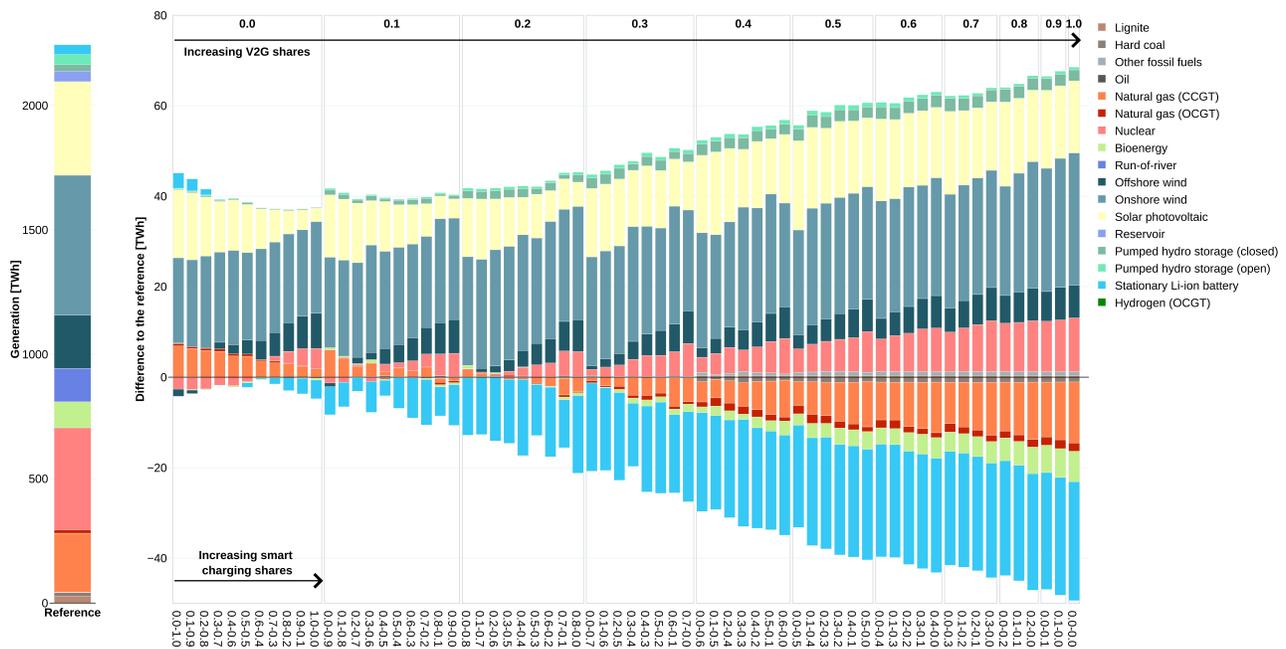}
    \caption{
    \textbf{Yearly electricity generation by technology for the reference (leftmost bar) and differences to the reference for all scenarios.}
    Panels displaying scenario results are sorted by increasing V2G shares (0\% in the leftmost panel, 100\% in the rightmost panel). Within each panel, scenarios are sorted by increasing smart charging shares. To simplify the reading, we remove the share of bidirectional charging from the scenario label. Hence, the first element of the x-axis label represents the share of smartly charging BEVs, while the second refers to the share of inflexibly charging BEVs.}
    \label{fig:dispatch}
\end{figure}

When there are no bidirectionally charging cars, the introduction of BEVs goes along with an increase in renewable generation which largely corresponds to the additional load consumed by the vehicles (Figure~\ref{fig:dispatch}, leftmost panel). This explains why the renewable share hardly changes in this case. If the V2G share increases, vRES generation increases beyond what is needed to cover the electricity demand of BEVs. This is because growing shares of bidirectionally charging vehicles increasingly enable to substitute a part of the power generation from combined-cycle gas turbines (CCGT), i.e. turbines with high variable costs, with wind and solar power. Generation from nuclear power plants also increases. In contrast, the use of stationary Li-ion batteries substantially decreases. To a smaller degree, this is also true for bioenergy. In other words, bidirectionally charging BEVs displace other flexibility options in the power sector. 

The impacts of BEVs on the optimal generation capacity portfolio are mostly in line with the dispatch effects (see Figure~\ref{fig:capacity}). In order to serve the additional load of a fully inflexible BEV fleet (leftmost bar of the leftmost panel), additional PV and wind onshore capacities have to be built, increasing overall system costs. In an fleet with more smartly charging cars but without V2G (moving right within the same panel), the need for additional PV and wind capacities decreases, as smart charging allows for a better use of capacities that are already built in the reference. Accordingly, renewable curtailment decreases when more vehicles charge smartly (Figure~\ref{fig:curtailment}). The main effect of growing shares of bidirectionally charging BEVs is that they increasingly allow to substitute stationary batteries and gas-fired peaker plants (OCGT), which smart charging cannot do. 

\subsection{Vehicle-to-grid helps cover the residual load}  

\begin{figure}[!ht]
    \centering
    \includegraphics[width=0.48\linewidth]{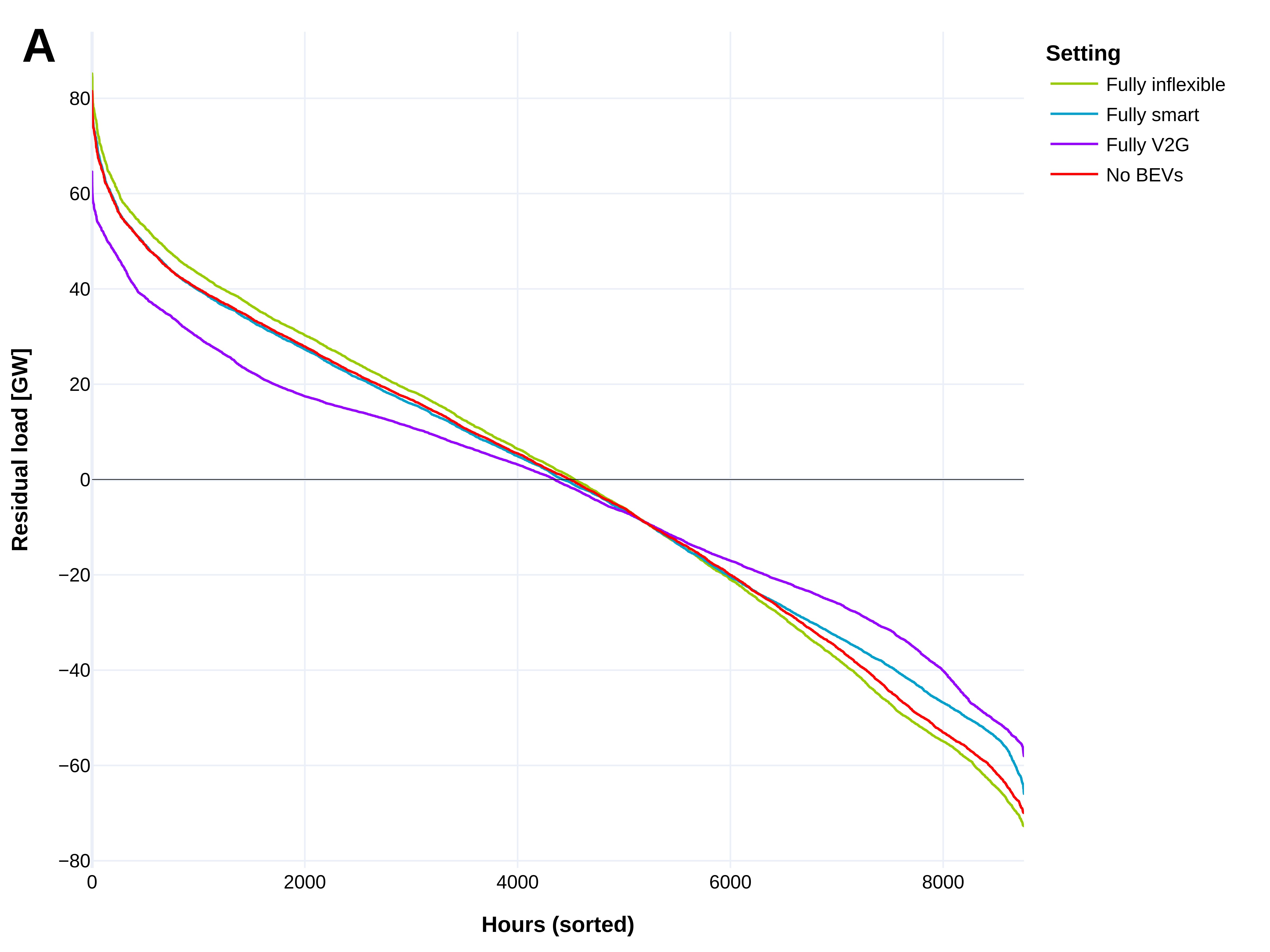}
    \includegraphics[width=0.48\linewidth]{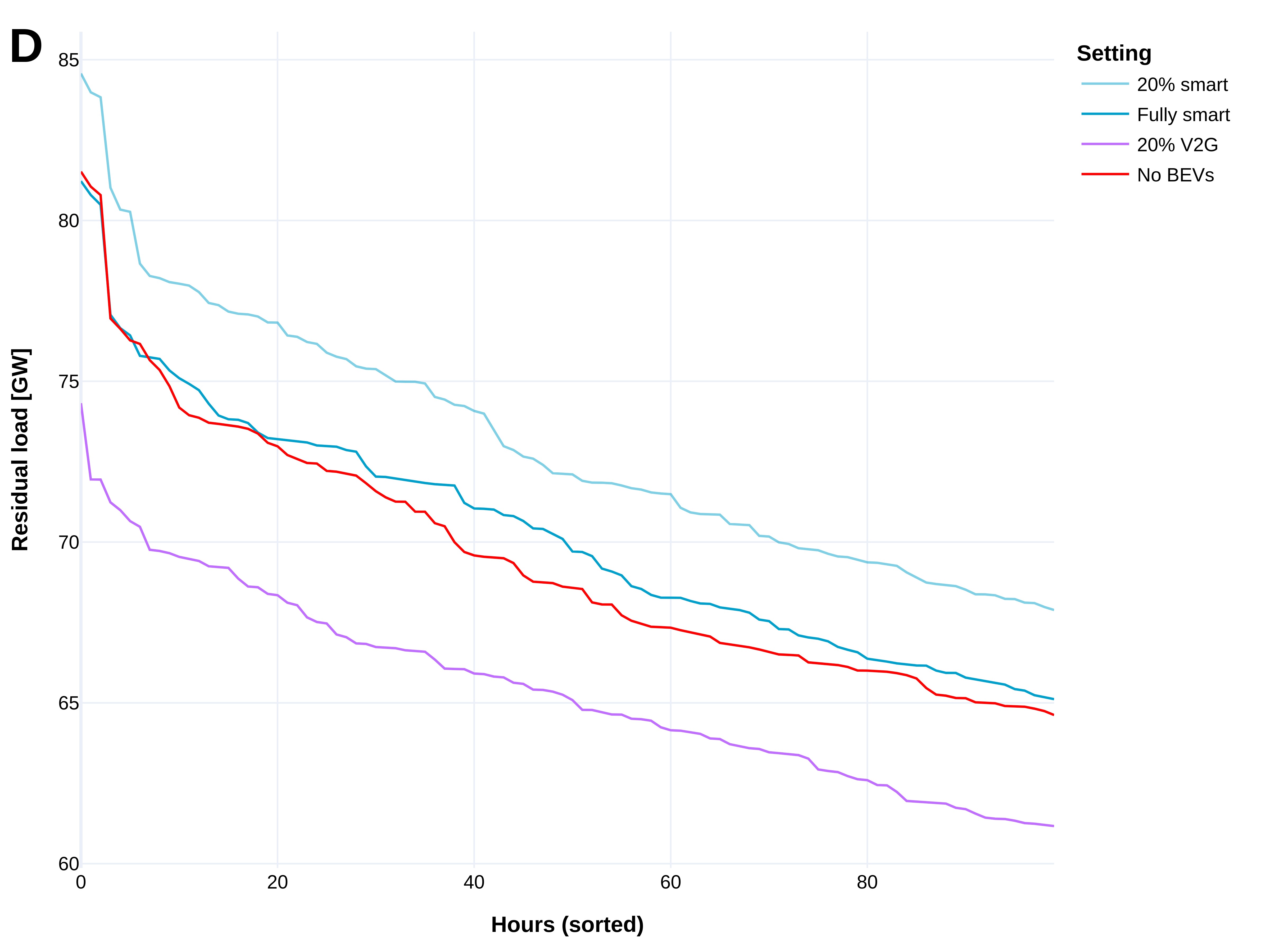}
    \includegraphics[width=0.48\linewidth]{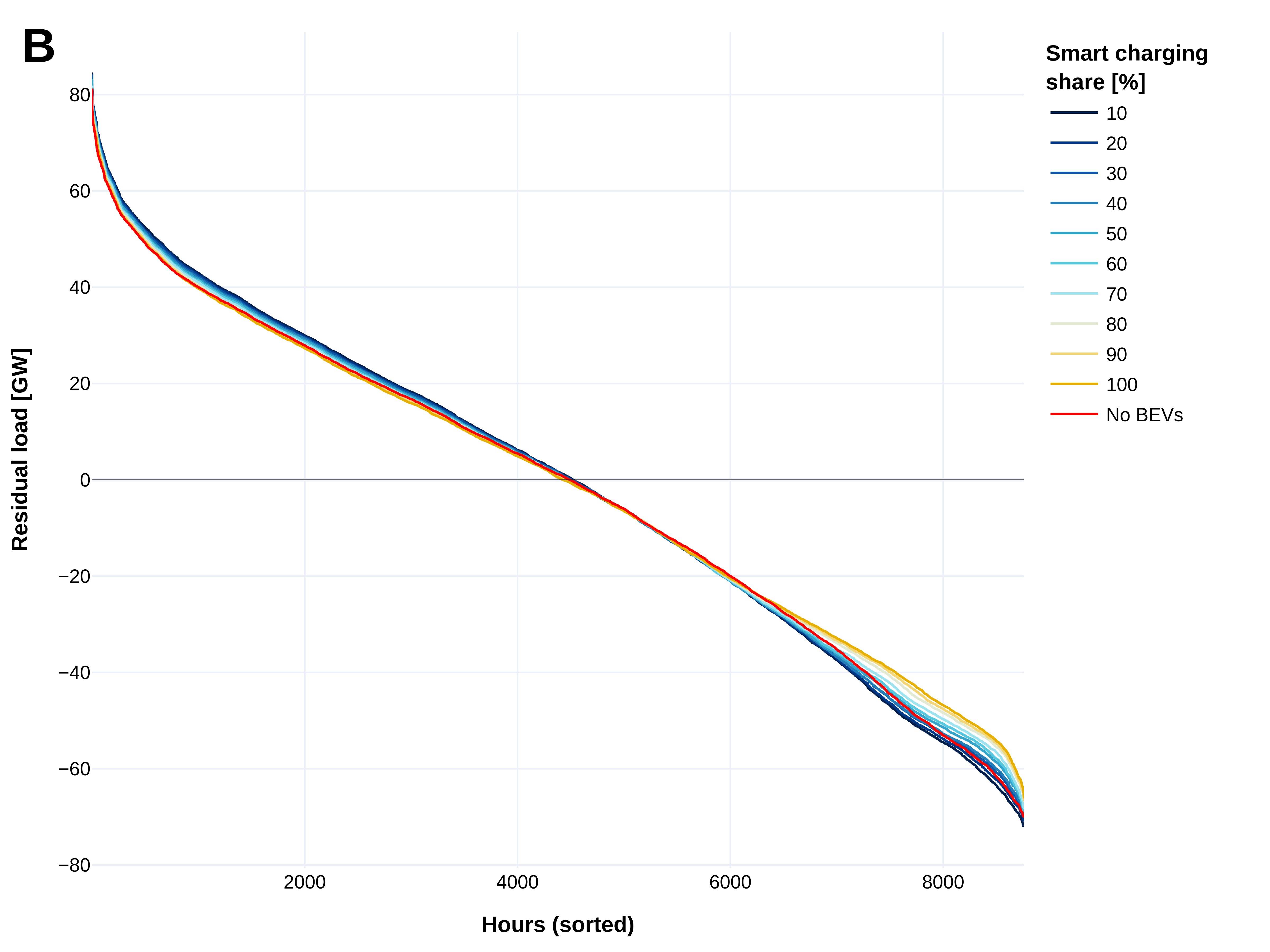}
    \includegraphics[width=0.48\linewidth]{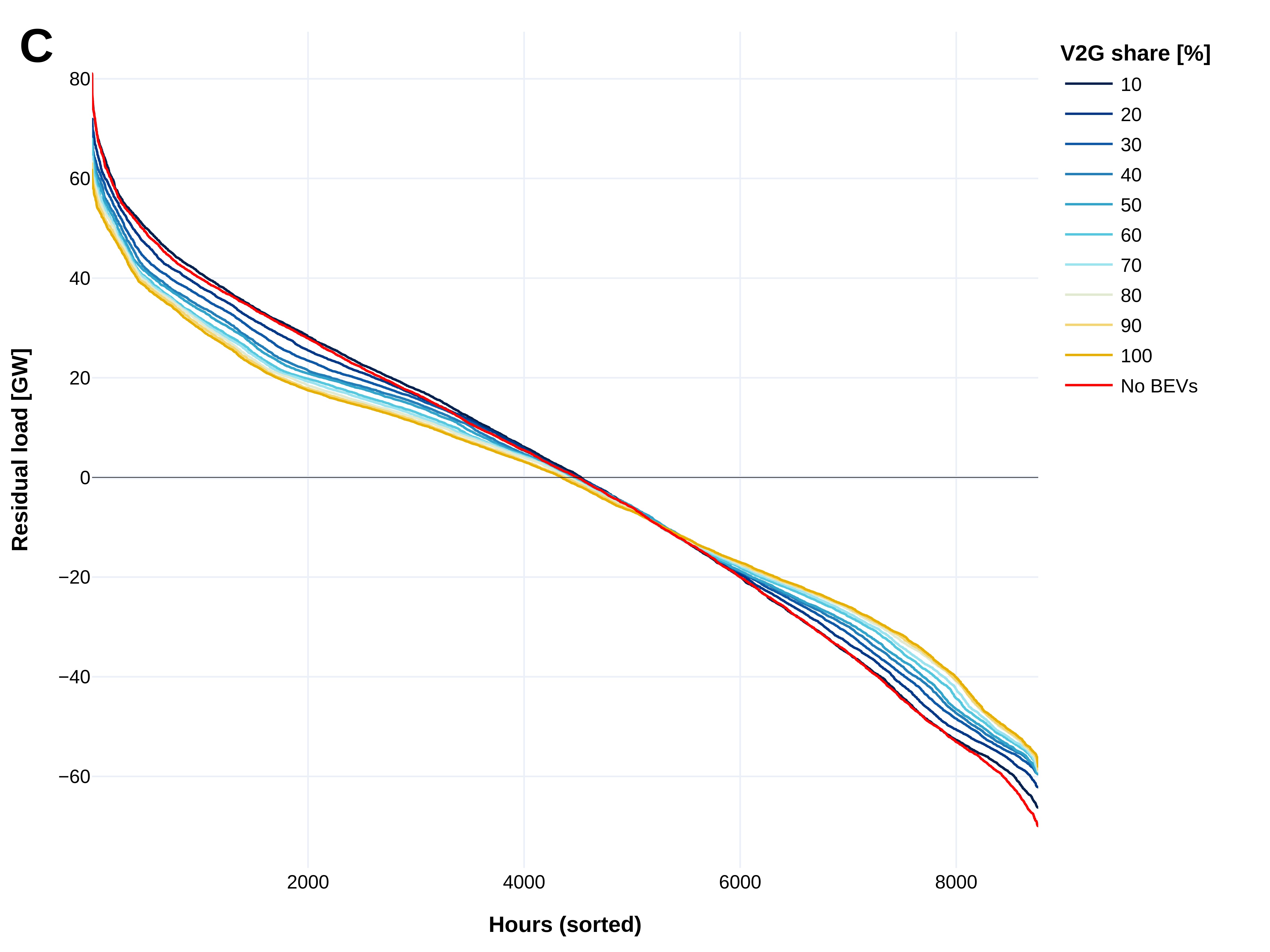}
    \caption{
    \textbf{Residual load duration curves (RLDC) for Germany for the reference and a selection of scenarios.}
    The reference (``No BEVs'') is depicted in all panels. The curves show residual load after curtailment and BEV charging and discharging. Counterclockwise starting from panel A, selected scenarios include: corner scenarios (fully inflexible/fully smart charging/fully V2G) (\textbf{A}); increasing smart charging shares and no V2G (\textbf{B}); increasing V2G shares and no smart charging (\textbf{C}); first 100 sorted hours for 20\% smart charging and no V2G, 100\% smart charging and for 20\% V2G and no smart charging (\textbf{D}). 
    }
    \label{fig:rldc}
\end{figure}

The introduction of a fully inflexible BEV fleet does not only increase the overall power demand but also the peak load, as indicated by the difference between the red and the light green residual load duration curves (RLDC) on the left-hand side of Figure~\ref{fig:rldc}, panel A. At the same time, the additional vRES capacities built to cover the additional demand also increase the renewable surplus energy (negative residual load) on the right-hand side of the RLDC. Smartly charging cars (blue line) can mitigate these effects to some extent, and V2G cars (purple line) even more so.

Growing shares of smartly charging BEVs increasingly allow to avoid charging at hours when the residual load is high (left-hand side of the RLDC in Figure~\ref{fig:rldc}, panel B). When the entire fleet is charging smartly, the positive part of the residual load curve is very similar to the one of the reference without vehicles (``No BEVs''). Besides, smart charging allows to make a better use of renewable surplus generation by charging during hours when the residual load is negative, increasing the RLDC on the right-hand side.

In comparison to smart charging, the flexibility potential of bidirectional charging is much larger, which leads to larger effects on both the left- and the right-hand sides of the curve (Figure~\ref{fig:rldc}, panel C). On the right-hand side, the increased charging of BEVs substantially pushes the curve upwards, as vehicles not only charge to serve their mobility needs, but also for later discharging to the grid. During hours of high residual load, i.e., on the left-hand side, bidirectionally charging cars allow not only to defer charging to more favorable hours, but also to serve the remaining load by discharging previously stored electricity. This effect is already sizable with only 20\% of the BEV fleet charging bidirectionally (see Figure~\ref{fig:rldc}, panel D). In contrast, smart charging, even when fully deployed in the BEV fleet, can only defer BEV charging, with a limited effect on the left-hand side of the RLDC.

\begin{figure}[!ht]
    \centering
    \includegraphics[width=1\linewidth]{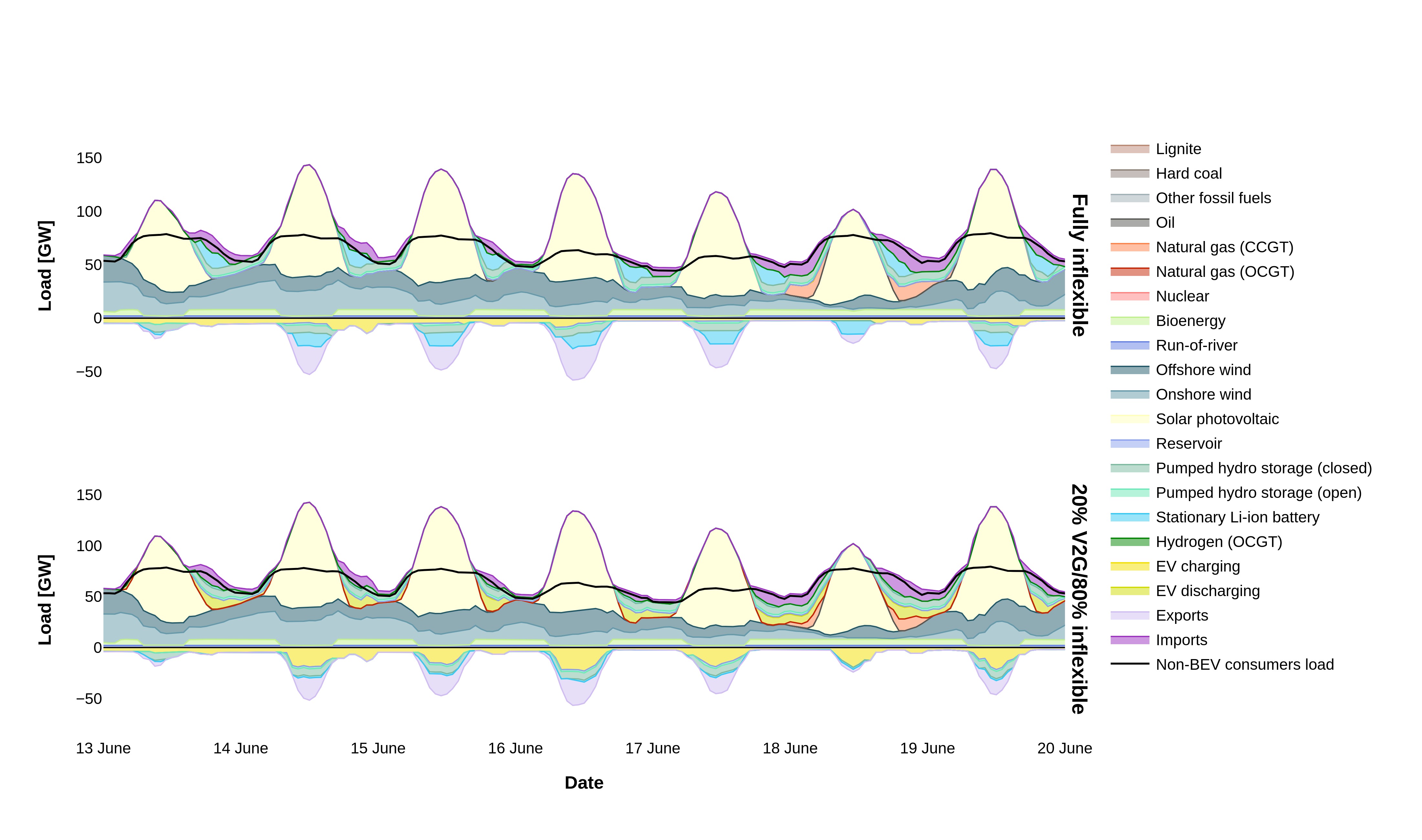}
    \caption{
    \textbf{Power generation by technology in Germany at an hourly resolution for a subset of summer days.}
    \textit{Upper panel --} Corner scenario with a fully inflexible BEV fleet. 
    \textit{Lower panel --} Scenario with 20\% of the BEV fleet charging bidirectionally and the remaining 80\% charging inflexibly. The load of non-BEV consumers does not include the electricity demand for electrolyzers.
    Negative (resp. positive) values for storage technologies represent charging (resp. discharging) episodes. 
   }
    \label{fig:dispatch_germany_spring}
\end{figure}

As depicted in Figure~\ref{fig:dispatch_germany_spring} for a fleet with 20\% bidirectional charging and a week in summer, BEV discharging replaces the discharging of stationary batteries. V2G cars typically charge at daytime and discharge at nighttime to smooth daily variations in solar PV generation. In winter, when the residual load is more shaped by wind power than by solar PV, the patterns of BEV charging and discharging are less regular (Figure~\ref{fig:dispatch_germany_winter}).

\subsection{Electricity bills and electricity prices show uneven distribution of system costs benefits}

\begin{figure}[!h]
    \centering
    \includegraphics[width=0.48\linewidth]{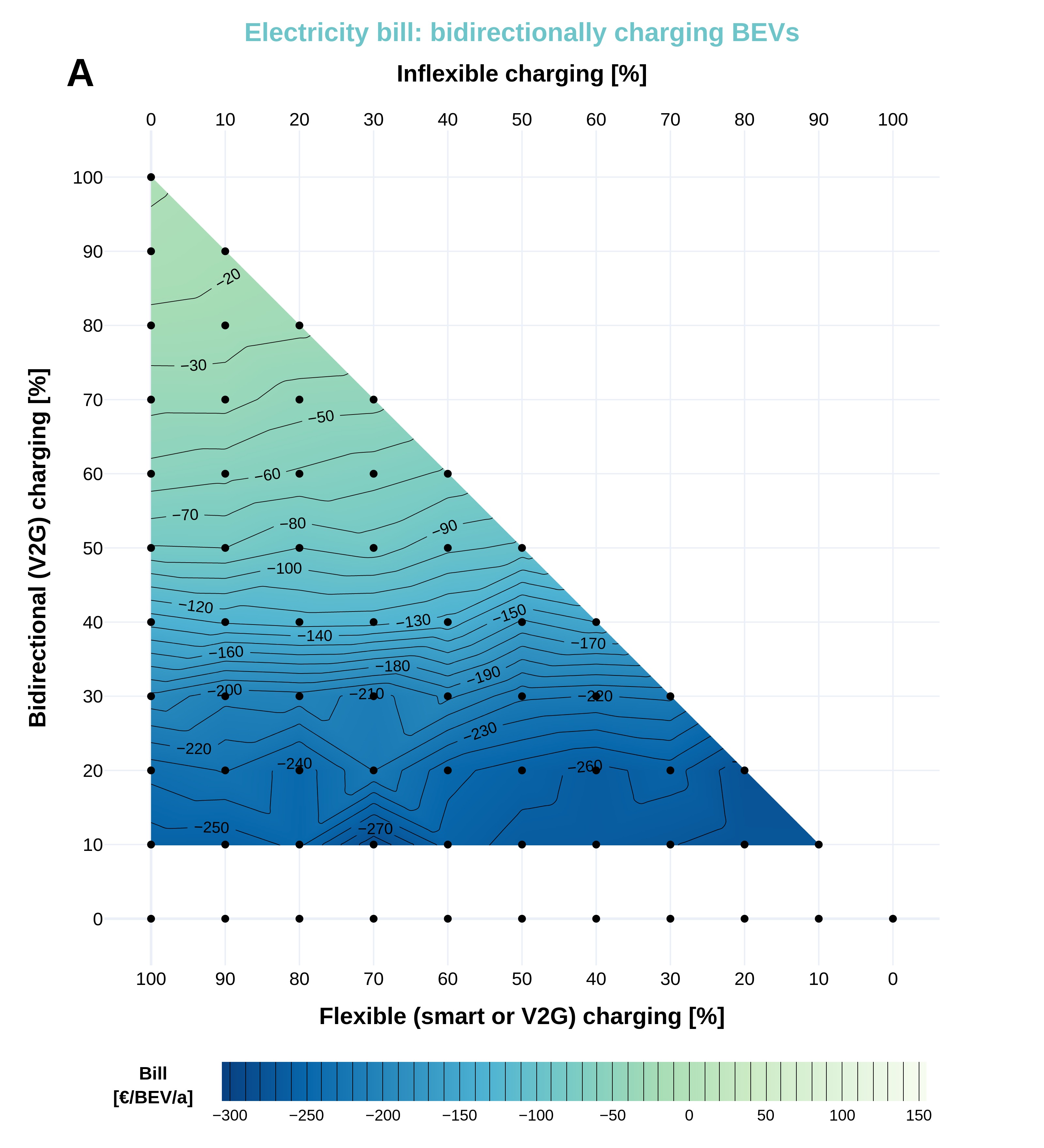}
    \includegraphics[width=0.48\linewidth]{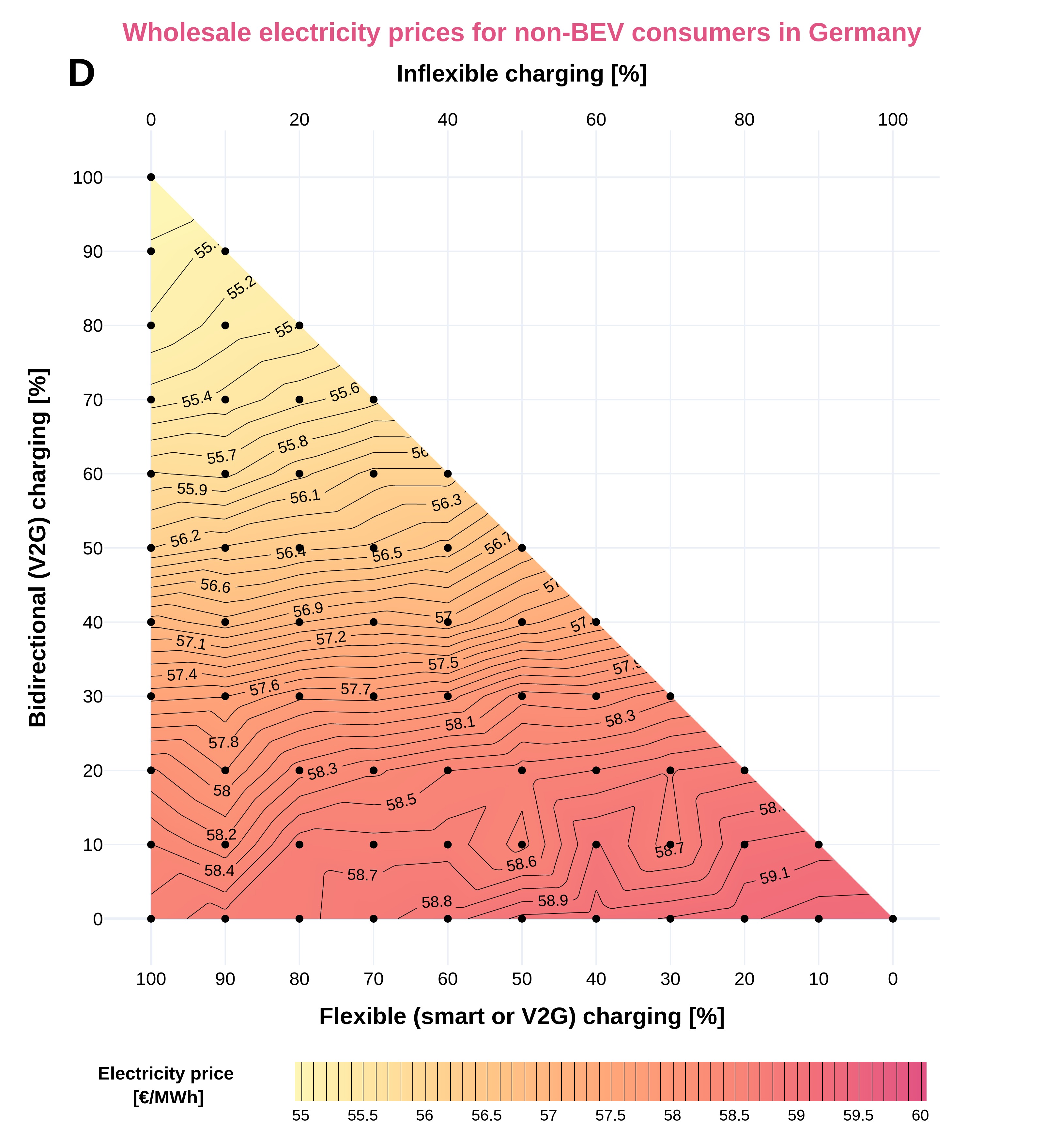}
    \includegraphics[width=0.48\linewidth]{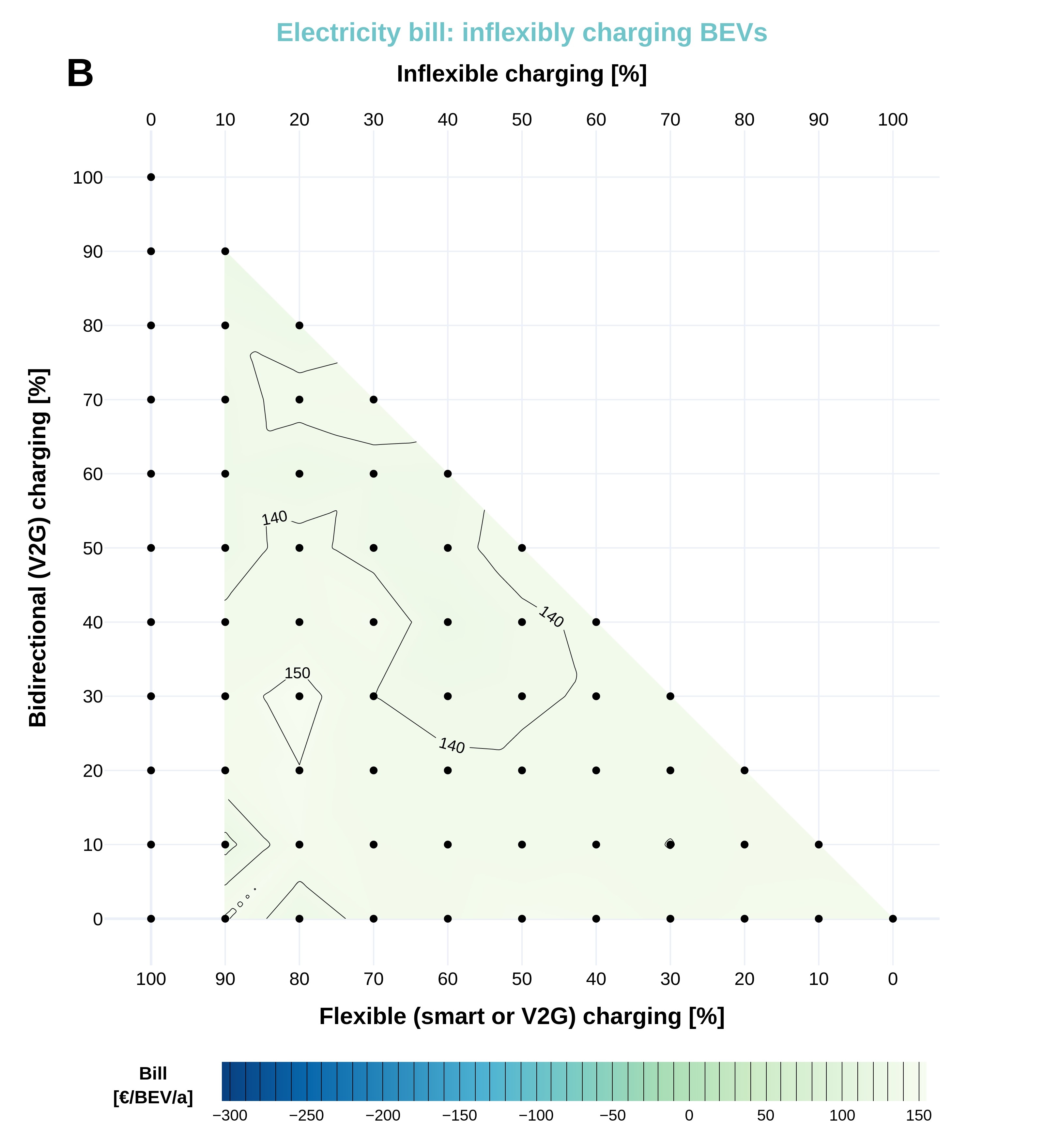}
    \includegraphics[width=0.48\linewidth]{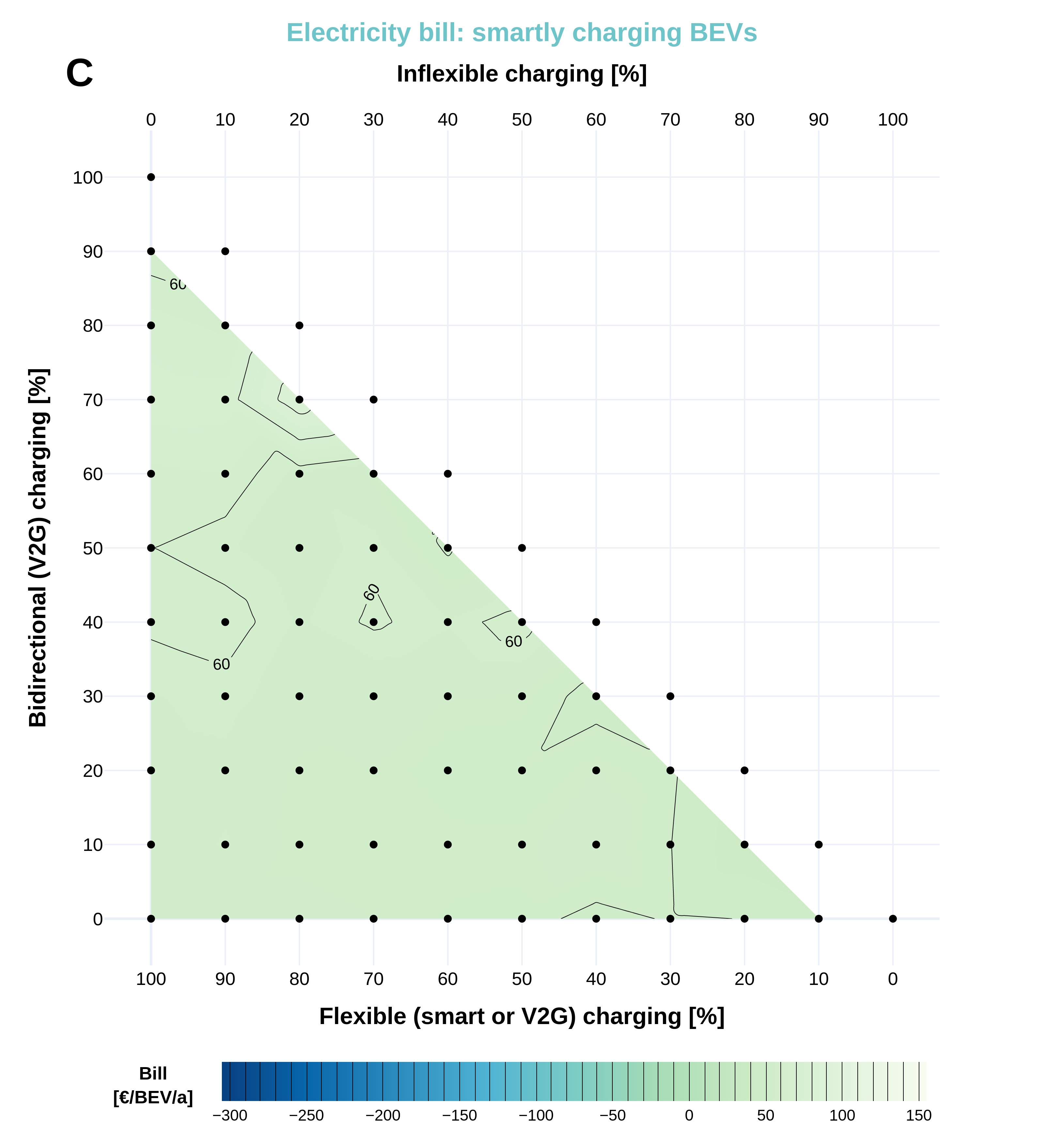}
    \caption{
    \textbf{Average yearly electricity bills for BEVs with different charging strategies (A-C) and average wholesale electricity prices for non-BEV consumers in Germany (D).}
    Counterclockwise, starting from panel A: bidirectionally charging BEVs (\textbf{A}); inflexibly charging BEVs (\textbf{B}); smartly charging BEVs (\textbf{C}); average wholesale prices for other consumers in Germany, excluding BEVs and electrolyzers.  Wholesale prices are weighted by the hourly non-BEV consumer load (\textbf{D}).
    }
    \label{fig:bills}
\end{figure}

Further analyses of model results show that the overall system costs savings of bidirectional charging are unevenly distributed across different types of electricity consumers. Interpreting the dual of the model's hourly energy balance as the wholesale price\autocite{brown2021decreasing}, we compute the average yearly amount paid for electricity consumed by BEVs, differentiated by charging strategy and by other electricity consumers. We find that electricity bills differ substantially between charging strategies (Figure~\ref{fig:bills}). Bidirectionally charging BEVs have by far the lowest bills (panel A). They are even negative, as arbitrage revenues from selling electricity back to the market are also factored in and outweigh charging costs. In other words, bidirectionally charging BEVs on average not only get their charging electricity for free, but even make arbitrage profits on the wholesale market. These profits can amount to nearly 300~euros per BEV per year at low shares of V2G and decrease with higher V2G penetration, as the prices decrease in periods of vehicle discharging and increase in periods of charging (Figure~\ref{fig:price-duration-curves}, panels B-D). However, V2G bills remain negative even if the entire BEV fleet is charging bidirectionally.

In contrast, inflexibly charging BEVs have positive yearly electricity bills of around 140~euros per BEV and year, and hardly benefit from higher V2G shares in the fleet (Figure~\ref{fig:bills}, panel B). Smartly charging BEVs have lower bills, between a third and half of what inflexibly charging BEVs pay (panel C), and bills tend to slightly increase with higher V2G shares. This is driven by the fact that smartly charging BEVs mostly charge in negative residual load hours, during which wholesale prices increase when the share of V2G grows, since bidirectionally charging vehicles strongly charge in these periods, beyond their mobility needs. In other words, here we find an instance of a superior flexibility option (bidirectional charging) cannibalizing the benefits of an inferior flexibility option (unidirectional charging).   

For other (non-BEV) electricity consumers in Germany, we do not show electricity bills, but weighted average wholesale electricity prices (Figure~\ref{fig:bills}, panel D). We find that these other consumers, which are assumed to have an inflexible demand, also benefit from increasing shares of V2G, indicated by lower average prices. Prices decrease from around 59~euros per MWh without any V2G to around 56~euros per MWh when half of the fleet charges bidirectionally. This reduction is largely driven by substantially lower peak prices (Figure~\ref{fig:price-duration-curves}, panel B) as well as reduced prices in hours when otherwise CCGT would set the price (Figure~\ref{fig:price-duration-curves}, panel C), enabled by V2G discharging. While the effect on average prices might appear small, it applies to a very large demand of 583~TWh. This implies that V2G benefits also spill over to other consumers. To some extent, this also applies to non-BEV electricity consumption in other countries (Figure~\ref{fig:heatmap_mean_baseload_price_wo_de}).

\begin{figure}[!ht]
    \centering
    \includegraphics[width=0.8\linewidth]{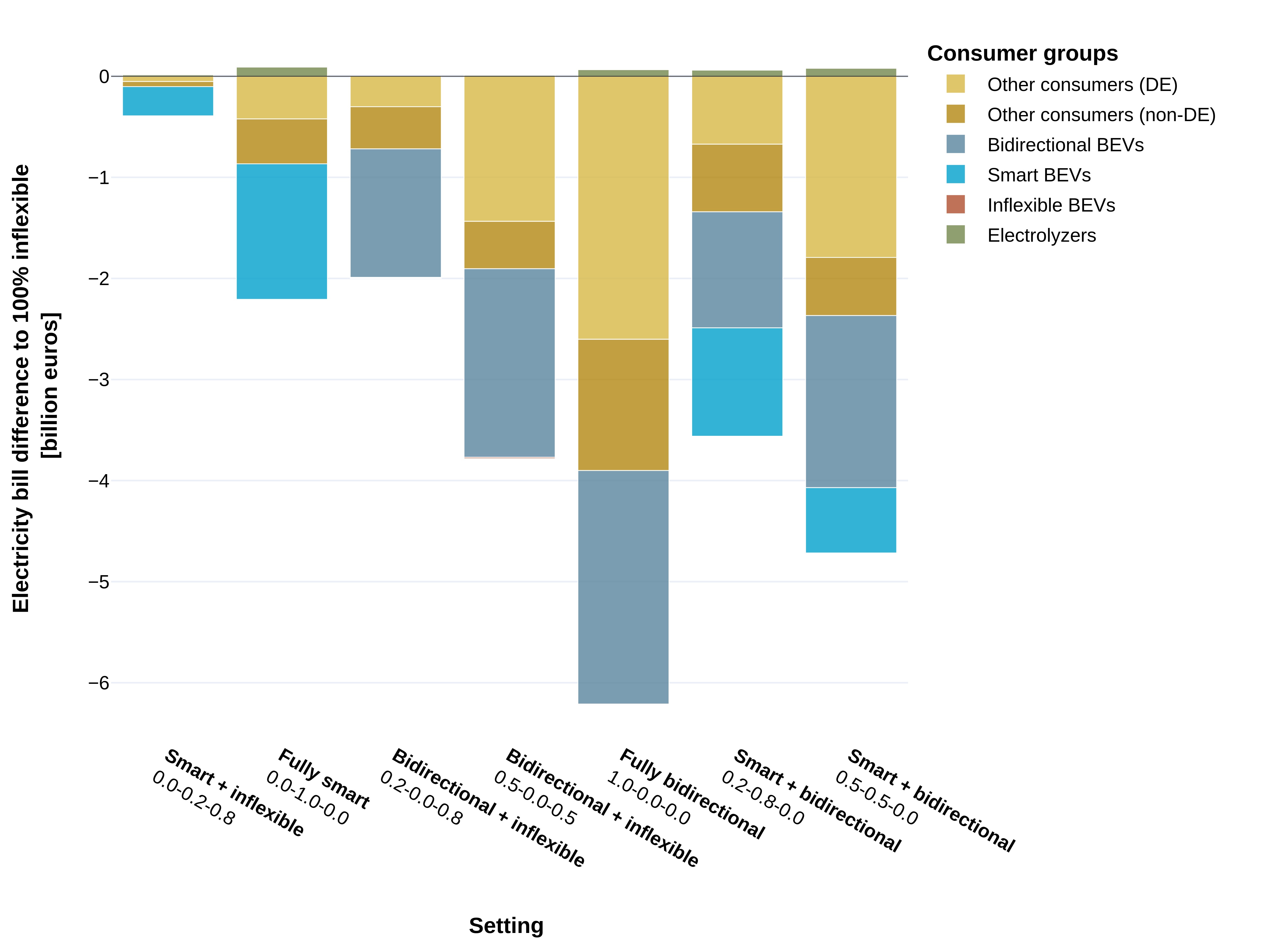}
    \caption{
    \textbf{Aggregate electricity bill differences compared to the 100\% inflexible corner scenario for selected cases and consumer groups.}
    Bill differences are decomposed for different consumer groups, namely BEV consumers (by charging strategy), electrolyzers and other consumers (Germany [DE] and other countries [non-DE]). For each group, the difference is calculated relative to the bill that this group would have to pay in the 100\% inflexible setting. For instance, if V2G cars represent 20\% of the BEV fleet, their bill in the 100\% inflexible setting is computed as 20\% of the aggregate bill paid by BEVs.
    }
    \label{fig:bill_distribution}
\end{figure}

To further illustrate the effects of BEV flexibility on different consumer groups, Figure~\ref{fig:bill_distribution} shows for selected cases how the aggregate electricity bill of each group changes when part of the fleet switches to more flexible operations, compared to a setting where the whole fleet charges inflexibly. It illustrates how the benefits of more flexible BEVs increasingly spill over to other consumers. While this is already true for increasing shares of smart charging (first two columns), this effect is even stronger for bidirectional charging. If 20\% of the fleet switches to V2G and the remaining 80\% continues to charge inflexibly, V2G cars see a substantial reduction in their aggregate electricity bills, and other consumers in Germany and in neighboring countries benefit only to a smaller extent (third column, ``0.2-0.0-0.8''). If the share of V2G increases to 50\%, not only the aggregate bill savings of V2G cars increase, but even more so those of non-BEV consumers (fourth column, ``0.5-0.0-0.5''). We find a qualitatively similar effect if more BEVs switch to V2G in a fleet of cars that are already smartly charging (two last columns). If the whole fleet switches to bidirectional charging (``Fully bidirectional''), the bills of other consumers in Germany decrease more (-2.6~billion euros) than the ones of the cars switching to V2G (-2.3~billion euros), while consumers abroad also benefit substantially (-1.3~billion euros). This is because other consumers increasingly benefit from lower wholesale prices in periods of V2G discharging, while growing fleets of bidirectionally charging cars self-cannibalize their arbitrage profits. The electricity bills of electrolyzers slightly increase, as they are competing with BEVs for cheap electricity.

\section{Discussion}
\label{sec:discussion}

Our results for a 2030 scenario with 15 million BEVs in Germany complement previous findings in the literature, which have been derived for other scenarios or geographic settings with different, and often less detailed, methods. For example, our analysis confirms the benefits of bidirectional charging\autocite{brown2018synergies, lund2008integration}. However, many previous analyses focus on extreme cases where the full vehicle fleet charges with a particular strategy\autocite{reibsch2024low, strobel2022joint, wei2022planning, crozier2020opportunity, szinai2020reduced, hanemann2017effects, forrest2016charging, loisel2014large, kiviluoma2011methodology}, or on a few specific mixes of charging strategies in the fleet \autocite{frank2025potential, syla2025assessing, bogdanov2024role, kamana2024driving, mangipinto2022impact, yao2022economic, lauvergne2022integration, wu2021benefits, taljegard2019impact, brown2018synergies}. Going beyond these studies, our systematic investigation of charging strategies shows that 30\% bidirectionally charging BEVs, even when combined with 70\% inflexibly charging cars, leads to lower costs than an entirely smartly charging fleet. If the share of bidirectional charging further increases to 50\%, system costs become even lower than in a setting without BEVs. We further extend the previous knowledge on the distributional effects of flexible BEV operations\autocite{emelianova2025welfare} by illustrating how other electricity consumers increasingly benefit from growing V2G shares in the fleet.

As in any modeling study, our numerical findings rely on a range of explicit or implicit assumptions. Some of these are likely to lead to an under- or overestimation of the benefits of flexible BEV charging and discharging. For others, the qualitative effects on results are not immediately clear.

On the one hand, we may underestimate the benefits of flexible BEV operations when compared to inflexible charging, as the latter already assumes relatively smooth charging profiles when vehicles are connected to the grid. Indeed, we use the ``immediate - balanced'' charging approach defined in \textit{emobpy}, which avoids extreme load peaks particularly in evening hours when many vehicles return to home at the same time. If the inflexible charging strategy was ``peakier'', for example if car batteries would be charged to the largest extent possible immediately after being connected to the grid\autocite{brown2018synergies, schill2015power}, the benefits of switching to a more flexible charging strategy would increase. Likewise, our assumption of a maximum 3.7~kW charging power rating at home constrains the flexibility potentials of BEVs when parked at home, compared to higher home charging power ratings of 11 or even 22~kW currently rolled out in many countries.

On the other hand, other assumptions likely lead to an overestimation of BEV flexibility and related power sector benefits. For example, we assume that the BEVs interact with the power system under perfect foresight. The benefits of flexible BEV operations modeled here can thus be interpreted as an upper boundary of what could be realized under limited foresight. Further, we do not factor in any additional costs for enabling the charging infrastructure or the cars themselves to be operated flexibly, which would reduce the cost savings they entail, particularly for V2G. It however appears plausible that such additional costs are low once these technologies are rolled out at scale. Likewise, we abstract from any inconvenience costs for car users which may reduce BEV flexibility benefits. While we assume that BEV owners face hourly wholesale market prices, either directly or via aggregators, BEV owners often face in reality time-invariant rates, which do not provide incentives for system-optimal (dis-)charging. We also neglect distribution or transmission grid constraints\autocite{reibsch2024low, strobel2022joint}, which might constrain BEV flexibility and lower respective system benefits. Our assumption that smart charging and even V2G are possible not only at home, but also when vehicles are plugged in at workplace and public charging stations might also overestimate the flexibility benefits of BEV, at least in a near-term 2030 scenario. For BEV fleets larger than 15 million, marginal flexibility benefits are likely to decrease due to an increasing temporal correlation between vehicle profiles. Similarly, considering a BEV rollout also in other neighboring countries might reduce the flexibility benefits of additional electric cars, at least if driving and charging patterns do not differ substantially across countries.

For other model assumptions, it is not immediately clear in which directions results might be distorted. For example, BEV flexibility effects may turn out to be larger or smaller in alternative weather years with different renewable generation patterns, even if previous work found only minor weather year effects on flexible BEV operations\autocite{gueret2024impacts}. Likewise, adding other sector coupling technologies such as a massive adoption of power-to-heat technologies might either increase or decrease the value of flexibility provided by BEVs, depending on how flexible the additional load is. 

Future research could focus on quantifying the effects of some of the aforementioned assumptions. For example, it would be of interest to investigate how results change under limited foresight, which might be approximated with more binding minimum and maximum BEV state of charge constraints in our current setting, or with a stochastic research design. Likewise, incorporating additional frictions or barriers to the rollout of bidirectional charging appears desirable. This could be done for instance by including infrastructure and inconvenience costs. However, these are not straightforward to quantify. Finally, assessing the role of flexible BEVs in massively sector-coupled continental renewable energy systems is computationally very demanding and may require to give up a very detailed modeling of BEV profiles to some degree. Investigating how fine-grained the BEV representation in highly sector-coupled power sector models should be would provide further valuable insights.

\section{Conclusion}

Our findings reveal that moderate shares of bidirectional charging (V2G) of less than 30\% can lead to lower system costs than a fully smartly charging BEV fleet in a central European 2030 scenario with 15 million BEVs in Germany. At a fleet share of 50\% and beyond, V2G even leads to overall system costs savings compared to a setting without BEVs. This implies that the \textit{flexibility effect} of V2G cars can outweigh the \textit{demand effect} of the whole BEV fleet, even if only half of the fleet charges bidirectionally and the other half charges inflexibly. This is not the case for unidirectional smart charging, including when all BEVs charge in this way. We further show that an improved system integration of low-cost variable renewable energy sources drives these cost effects. In particular, bidirectionally charging vehicles do not only help integrate renewable surplus energy, but also provide electricity in peak residual load hours. This proves to be particularly valuable as the need for stationary batteries and other expensive peak generation options decreases. We further show that V2G cars internalize a substantial share of the system benefits they entail, but with growing shares of V2G, these benefits increasingly spill over to other electricity consumers.

As the electrification of the car fleet is still at early stages in many countries, better insights on how charging strategies interact can help design incentives for consumers, industries and infrastructure planners to embrace V2G, which emerges as a game changer for the power system. Based on our results, we conclude that it is more important for policymakers to enable even a relatively small share of the electric vehicle fleet to engage in V2G rather than to incentivize every car to charge smartly. Policymakers should thus focus on facilitating the rollout of V2G, which requires to foster the deployment of the necessary charging infrastructure. In this respect, vehicle-to-home and vehicle-to-building could provide a good starting point for bidirectional charging, with a perspective to subsequently roll out a wider network of bidirectional charging points at other locations. At the same time, policymakers should enable vehicles to respond to wholesale market price signals and remove corresponding regulatory barriers.

\section{Methods}
\label{sec:methods}

The methodological framework combines two open-source tools. First, we generate electric vehicle time series with the probabilistic tool \textit{emobpy}\autocite{gaete-morales_open_2021}. We then use these time series in a linear cost-minimizing capacity expansion model of the power sector, calibrated to European data for a 2030 scenario. 

\subsection{Electric vehicle time series generation}

Our analysis is based on 100 BEV profiles. Each profile consists of three time series generated by the open-source probabilistic tool \textit{emobpy}\autocite{gaete-morales_open_2021} at a 15-minute time step and aggregated at an hourly resolution to match the time resolution of the power sector model. These time series are generated based on empirical distributions retrieved from publicly available data from a German national representative travel survey, \textit{Mobility in Germany}\autocite{eggs2018mobilitat}. These distributions are: number of trips per day, departure time by destination type and joint distribution of trip distance and trip duration. Destination types considered are workplace, errands, escort, leisure and home. More details on these distributions are provided in Tables 1-3 of the paper describing \textit{emobpy} in detail\autocite{gaete-morales_open_2021}. 

Depending on the charging strategy considered for each BEV profile, time series included in the power sector model differ. If a BEV profile is assumed to be inflexibly charging, a respective grid electricity demand time series is added to the exogenous electric load. Here, we choose the option ``immediate - balanced'' from \textit{emobpy}, which means that cars are charged with a flat profile during the entire time they are connected to the grid. Note that this avoids extreme load peaks which would occur were vehicles charged at the full available power rating immediately after connecting to the grid. For smartly or bidirectionally charging BEV profiles, two other time series are needed in the power sector model: a driving electricity consumption time series and a grid charging availability time series. They restrict the ability of the BEV to charge to specific time slots and power ratings that are stochastically generated by \textit{emobpy}. The power sector model can thus decide when and how much to charge (and, if V2G is assumed, discharge), provided that the vehicle is plugged to a charging station and ensuring that the battery is always full enough to undertake the next trips. The grid charging availability time series relies on assumptions on charging stations' power ratings and locations. In this study, charging is possible at home (3.7~kW), in the public area (22~kW) or at the workplace (11~kW).  

Each BEV profile is also characterized by parameters such as its motor power, battery energy capacity, heat transfer characteristics, geometry and weight. Based on data from the Electric Vehicle Database\autocite{electric_vehicles_database}, we use four different BEV models corresponding to Renault Zoe, Volkswagen ID.3, Tesla Model 3 and Hyundai Kona. Assumed characteristics are provided in Table 4 of the aforementioned paper\autocite{gaete-morales_open_2021}. Overall, we include 100 BEV profiles (20 Volkswagen ID.3, 23 Renault Zoe, 24 Hyundai Kona and 33 Tesla Model 3). 

\subsection{Power sector modeling} 

We use the open-source power sector model DIETER\autocite{zerrahn2017long, gaete-morales_open_2021} (\textit{Dispatch and Investment Electricity Tool with Endogenous Renewables}) in its julia version\footnote{Available on GitLab at: \href{https://gitlab.com/diw-evu/dieter_public/DIETERjl}{https://gitlab.com/diw-evu/dieter\_public/DIETERjl}.}. The model is a linear cost-minimizing problem with perfect foresight, which optimizes capacity and dispatch decisions for various generation and storage technologies. The hourly resolution for a representative year allows to account for the daily and seasonal variability of renewable energy sources. Model results can be interpreted as the outcomes of an idealized, frictionless market. Different versions of the model have been used to investigate policy-relevant questions related to the power sector coupling with battery electric vehicles\autocite{gaete2024power, gueret2024impacts}, residential electric heating\autocite{roth2024power, schill2020flexible} and green hydrogen\autocite{kirchem2023power, stoeckl2021optimal} and related to flexibility requirements in variable renewable energy systems\autocite{schmidt2025mix, kittel2024coping}. The geographical scope centers around Germany and includes nine of its interconnected neighbors (Austria, Belgium, Czech Republic, Denmark, France, Luxembourg, the Netherlands, Poland, Switzerland) and Italy. Input data generally reflects assumptions for the year 2030.

\subsubsection{Technologies and capacity constraints}

We consider twelve generation technologies, including renewables (bioenergy, run-of-river, solar photovoltaic, wind onshore and wind offshore) and conventional technologies such as lignite, hard coal, natural gas (OCGT and CCGT), nuclear, oil and a residual other fossil fuel technology. We also consider five storage technologies, namely Li-ion batteries, pumped-hydro (closed and open), hydro reservoirs and a long-duration storage technology, which consists in compressed green hydrogen stored in caverns and reconverted to electricity via H$_2$-ready gas turbines. The latter is assumed possible only in Denmark, France, Germany, the Netherlands and Poland due to limited cavern potentials in the other countries. 

Based on data from a Ten-Year Network Development Plan (TYNDP) of ENTSO-E\autocite{entsoe_tyndp_2018}, capacities for renewable technologies and nuclear are fixed, while their dispatch is endogenous and unconstrained. The only exception is Germany, where capacity expansion for wind onshore, wind offshore and solar photovoltaic is left unconstrained, with an upper capacity bound of 30 GW for wind offshore to ensure policy relevance. Besides, for all countries, the capacity of hydropower technologies are fixed to levels provided by the ENTSO-E Pan-European Climate Database (PECD 2021.3)\autocite{de2022entso} to reflect very limited expansion potentials for these technologies. For similar reasons, bioenergy capacities for all countries are fixed. Investment in conventional technologies is left free up to capacity bounds provided by the TYNDP, except for gas technologies in Germany, which are unconstrained. Country-specific capacity constraints for all considered technologies are detailed in Table \ref{tab:parameters_capacitybounds}. Free investments in renewable and gas technologies in Germany enable to respond to the additional BEV load. In other countries, fixed renewable capacities ensure policy relevance. Making the dispatchable conventional technologies endogenous, on the other hand, prevents overly high available capacities, and instead results in a long-run equilibrium level of firm capacity.

Interconnections across countries are based on net transfer capacities provided by ENTSO-E\autocite{entsoe_tyndp_2018} and detailed in Table \ref{tab:parameters_ntc}. 

\subsubsection{Weather and cost data}

Capacity factors time series for solar, wind onshore, wind offshore and inflow time series for run-of-river, reservoirs and open pumped-hydro are taken from the ENTSO-E Pan-European Climate Database (PECD 2021.3)\autocite{de2022entso} for the weather year 2008. 

Cost parameters are mostly based on data from the Danish Energy Agency\autocite{dea_techdata2024} and are detailed in Table~\ref{tab:parameters_generation} and Table~\ref{tab:parameters_storage}. Fossil generation technologies face a carbon price of 130~euros per tCO$_2$eq that adds up to their operational costs. Transmission and distribution network costs are not considered.

\subsubsection{Demand data}

Electricity demand data for all countries are taken from the ENTSO-E Pan-European Climate Database (PECD 2021.3)\autocite{de2022entso}, using estimates for 2030 and the weather year 2008. The German yearly load amounts to 583~TWh. The overall assumed load for each country is detailed in Table \ref{tab:parameters_load}. For Germany, we additionally include the electricity demand for BEVs in scenarios. The yearly driving consumption amounts to 33.2~TWh for a 15~million BEV fleet. This corresponds to an average of~2.2~MWh per BEV per year. Finally, we also include for Germany a 30~TWh demand for industrial green hydrogen, assumed to be evenly distributed across the year, that needs to be produced via electrolysis using renewable energy sources.

\section*{Author contributions}

Conceptualization: A.G. \& C.G.-M. \& W.-P.S.;
Methodology: C.G.-M. \& W.-P.S.;
Software: A.G. \& C.G.-M.;
Formal analysis: A.G. \& C.G.-M \& W.-P.S.; 
Investigation: A.G. \& C.G.-M. \& W.-P.S.; 
Data Curation:  A.G. \& C.G.-M.; 
Writing - Original Draft: A.G.;  
Writing - Review \& Editing: W.-P.S.; 
Visualization: A.G. \& C.G.-M.; 
Funding acquisition: W.-P.S. 

\section*{Data and code availability}

The code for data and results processing as well as all input data are available in a public GitLab repository available at \href{https://gitlab.com/diw-evu/projects/bev_flexibility/-/tree/newparam_feb25?ref_type=heads}{this link}. DIETERjl is publicly available on \href{https://gitlab.com/diw-evu/dieter_public/DIETERjl}{GitLab}. The python package for \textit{emobpy} is available at \href{https://pypi.org/project/emobpy/}{https://pypi.org/project/emobpy/} and a \href{https://emobpy.readthedocs.io/en/latest/index.html}{documentation page} provides further instructions. MiD data are also publicly available and can be retrieved at \href{https://www.mobilitaet-in-deutschland.de/archive/MiT2017.html}{https://www.mobilitaet-in-deutschland.de/archive/MiT2017.html}. 

\section*{Acknowledgments}
We thank the ``Transformation of the Energy Economy'' research group at the German Institute for Economic Research (DIW Berlin) for helping to maintain and develop DIETER as well as for fruitful scientific exchange. We thank the participants of a power sector modeling seminar at 50Hertz for their useful feedback. We acknowledge a research grant by the German Federal Ministry of Education and Research via the ``Ariadne'' projects (Fkz 03SFK5NO \& 03SFK5NO-2).

\printbibliography

\appendix
\setcounter{figure}{0}
\renewcommand{\thefigure}{SI.\arabic{figure}}
\setcounter{table}{0}
\renewcommand{\thetable}{SI.\arabic{table}}
\renewcommand{\thesubsection}{SI.\arabic{subsection}}

\newpage 

\section*{Supplemental Information}
\subsection{Additional graphs}

\begin{figure}[!ht]
    \centering
    \includegraphics[width=1.0\linewidth]{plot_capacity.jpeg}
    \caption{
    \textbf{Electricity generation and storage capacity mix by technology for the reference (leftmost bar) and differences to the reference for all scenarios.}
    Panels displaying scenario results are ordered by increasing V2G shares (0\% in the leftmost panel, 100\% in the rightmost panel). Within each panel, scenarios are ordered by increasing smart charging shares. To simplify the reading, we remove the share of bidirectional charging from the scenario label. Hence, the first element of the x-axis label refers to the share of smartly charging BEVs, the second to the share of inflexibly charging BEVs.
    }
    \label{fig:capacity}
\end{figure}

\begin{figure}[!ht]
    \centering
    \includegraphics[width=0.45\linewidth]{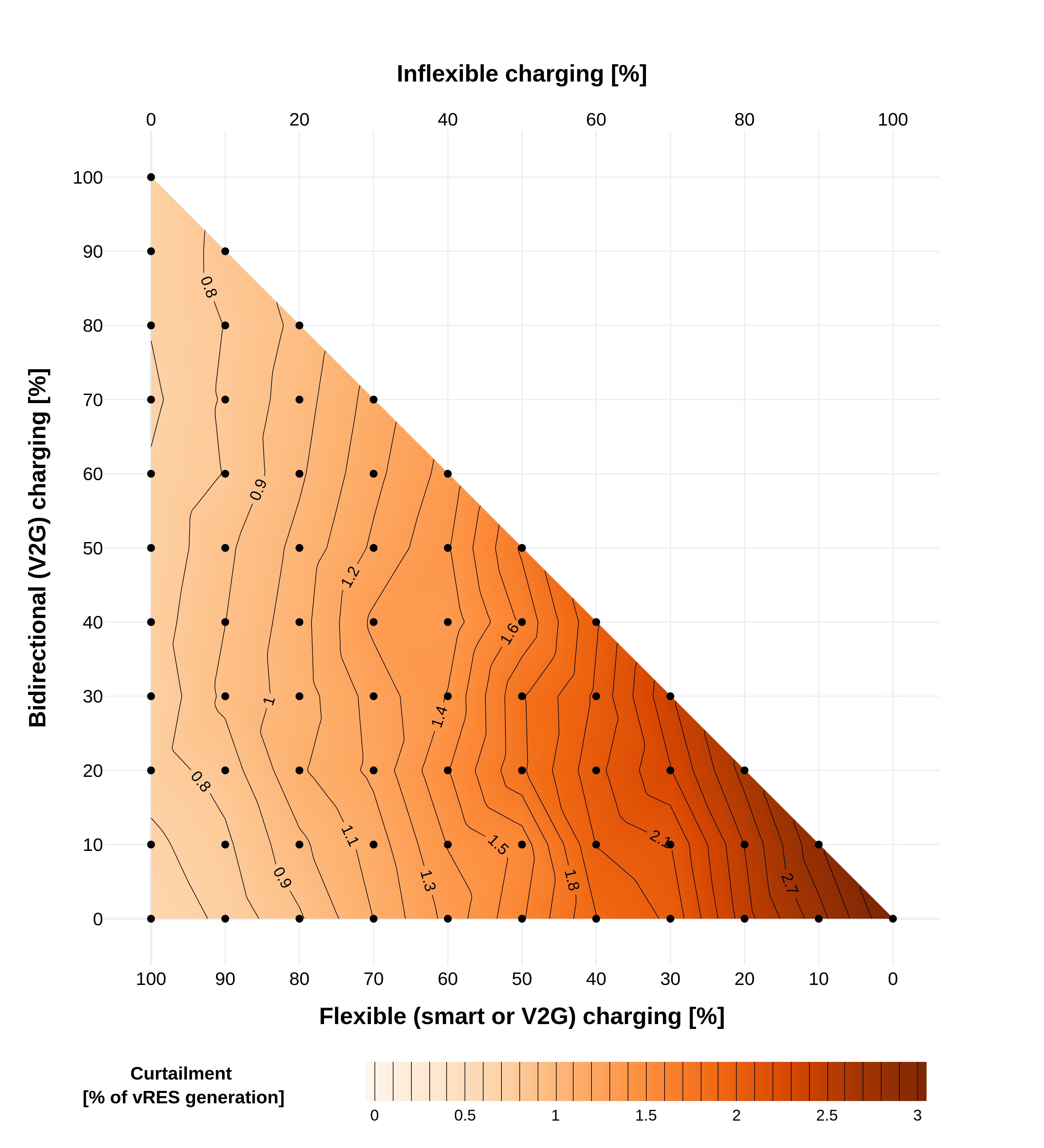}
    \caption{
    \textbf{Total curtailed electricity generation, expressed in percentage of the total vRES electricity generated before curtailment.}
    vRES technologies refer to run-of-river hydro, solar photovoltaic, wind offshore and wind onshore.}
    \label{fig:curtailment}
\end{figure}

\begin{figure}[!ht]
    \centering
    \includegraphics[width=0.9\linewidth]{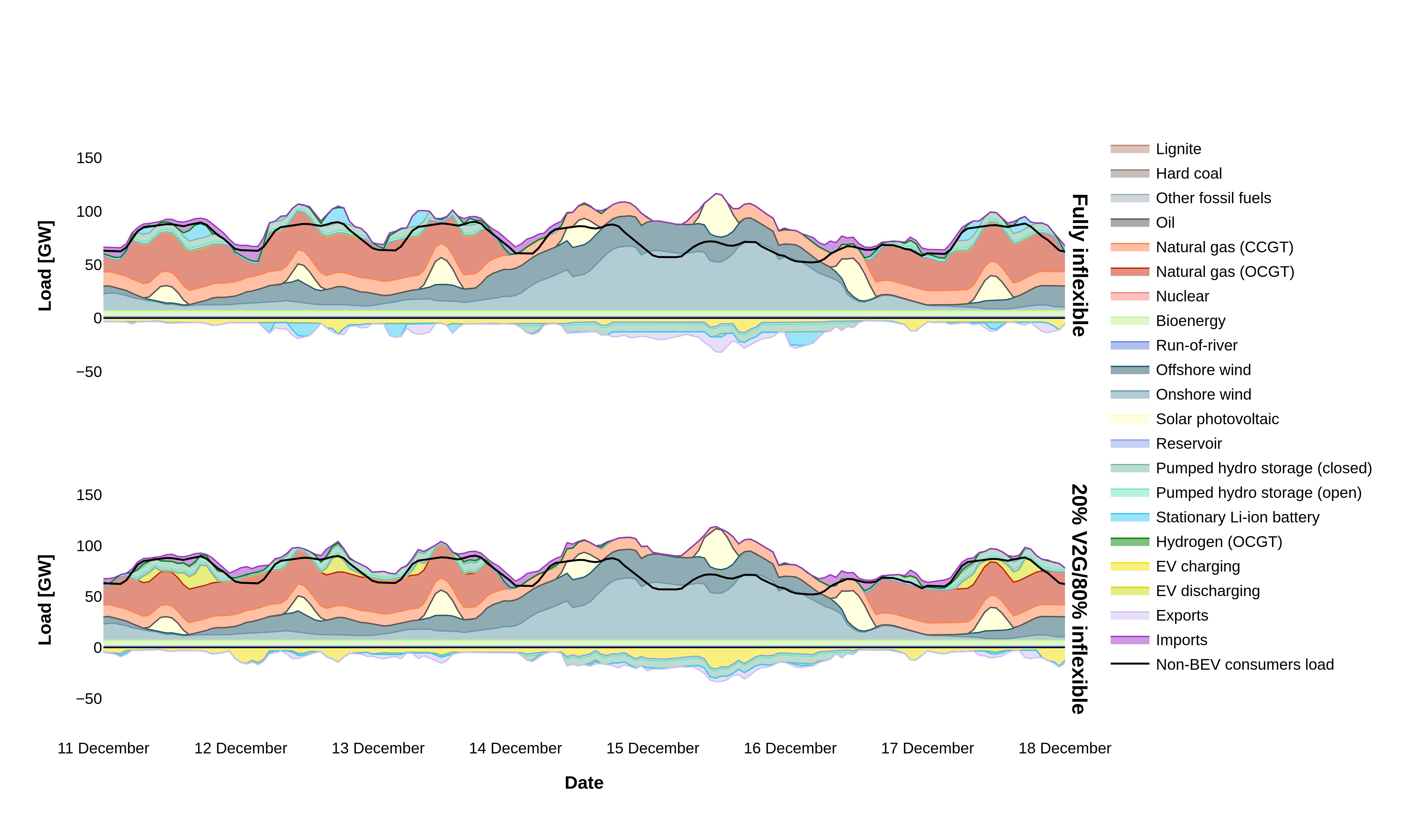}
    \caption{
    \textbf{Power generation by technology in Germany at an hourly resolution for a subset of winter days.}
    \textit{Upper panel --} Corner scenario with a fully inflexible BEV fleet. 
    \textit{Lower panel --} Scenario with 20\% of the BEV fleet charging bidirectionally and the remaining 80\% charging inflexibly. The load of non-BEV consumers does not include the electricity demand for electrolyzers.
    Negative (resp. positive) values for storage technologies represent charging (resp. discharging) episodes. 
    }
    \label{fig:dispatch_germany_winter}
\end{figure}

\begin{figure}[!ht]
    \centering
    \includegraphics[width=0.45\linewidth]{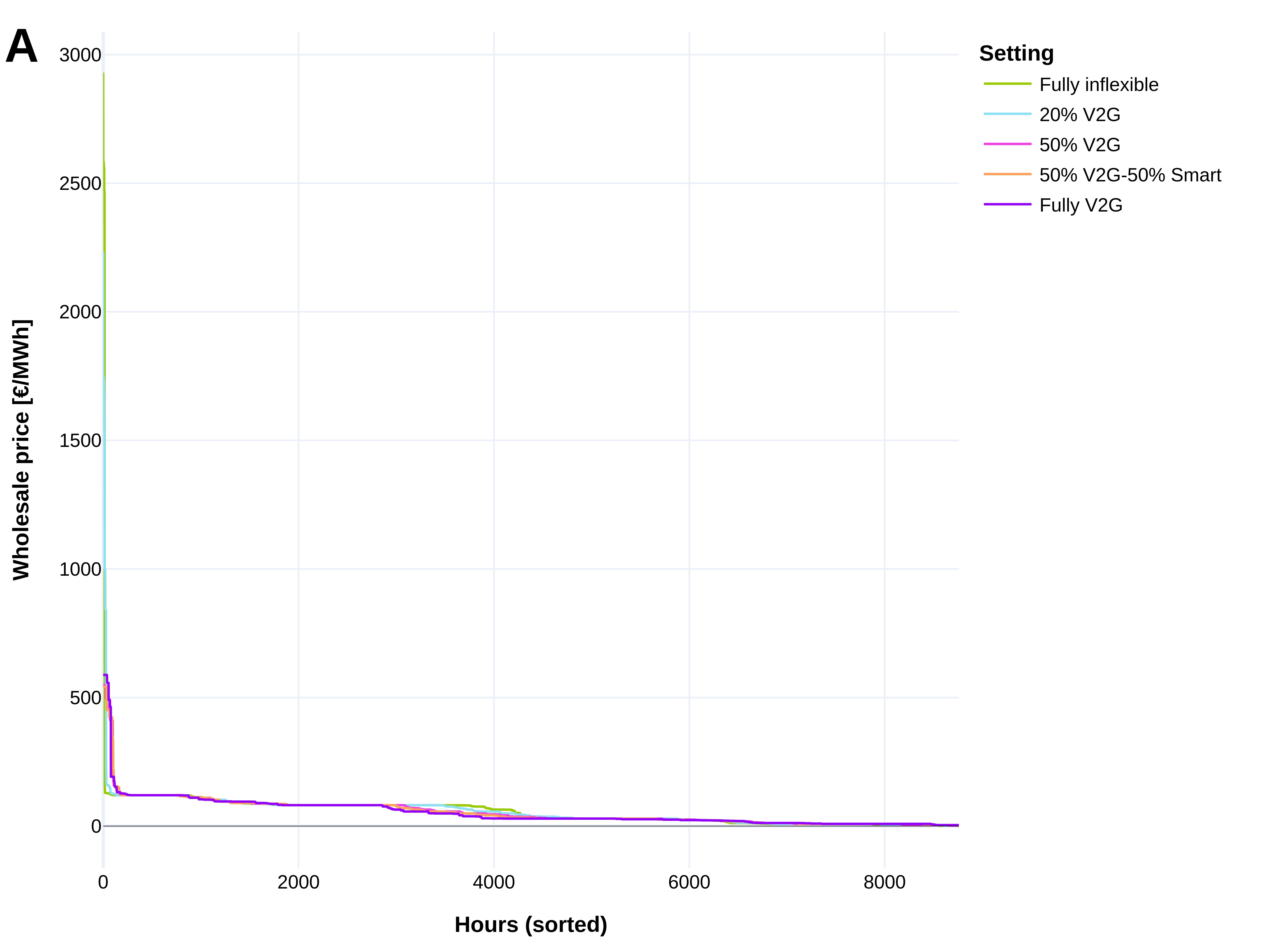}
    \includegraphics[width=0.45\linewidth]{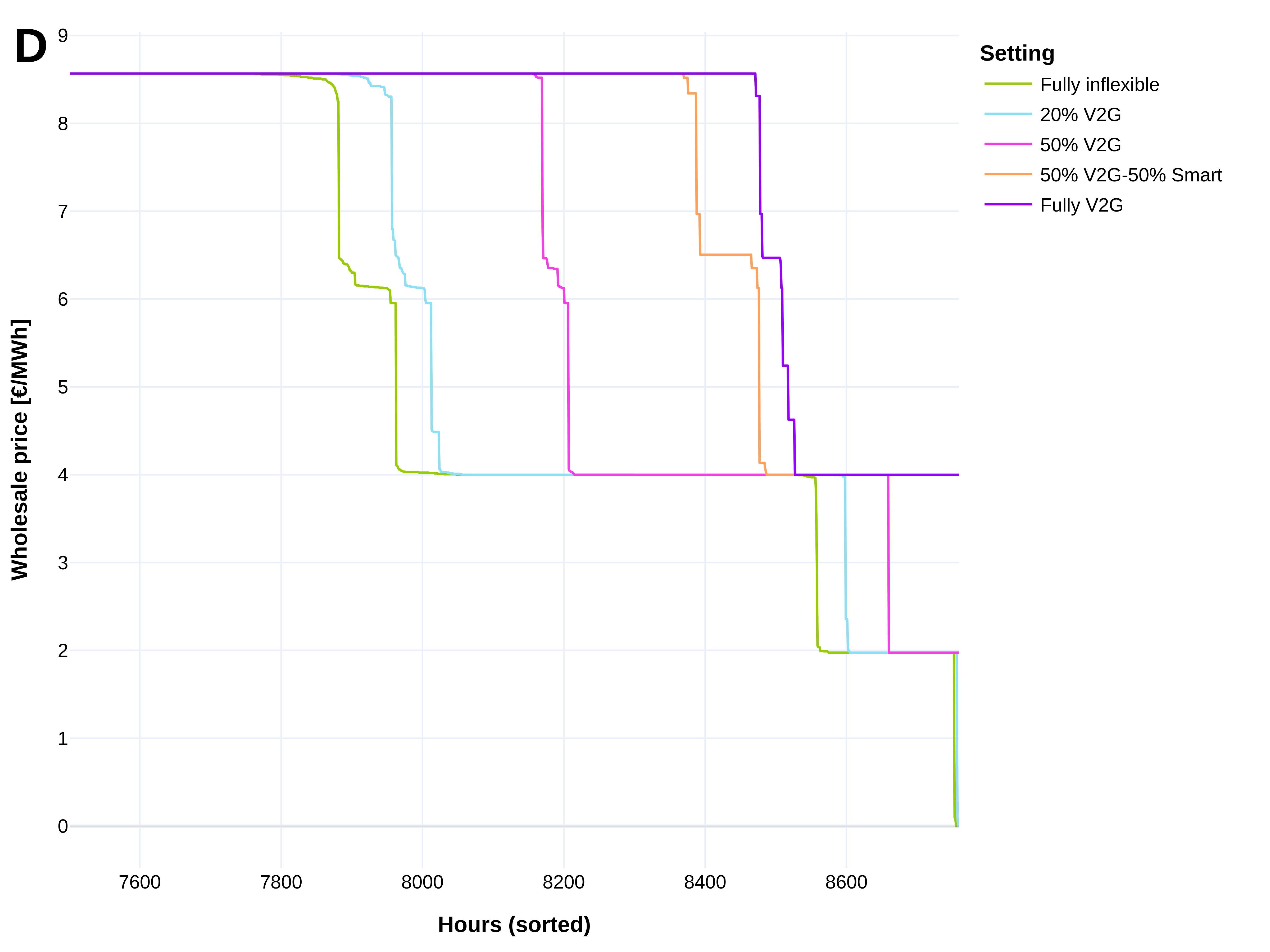}
    \includegraphics[width=0.45\linewidth]{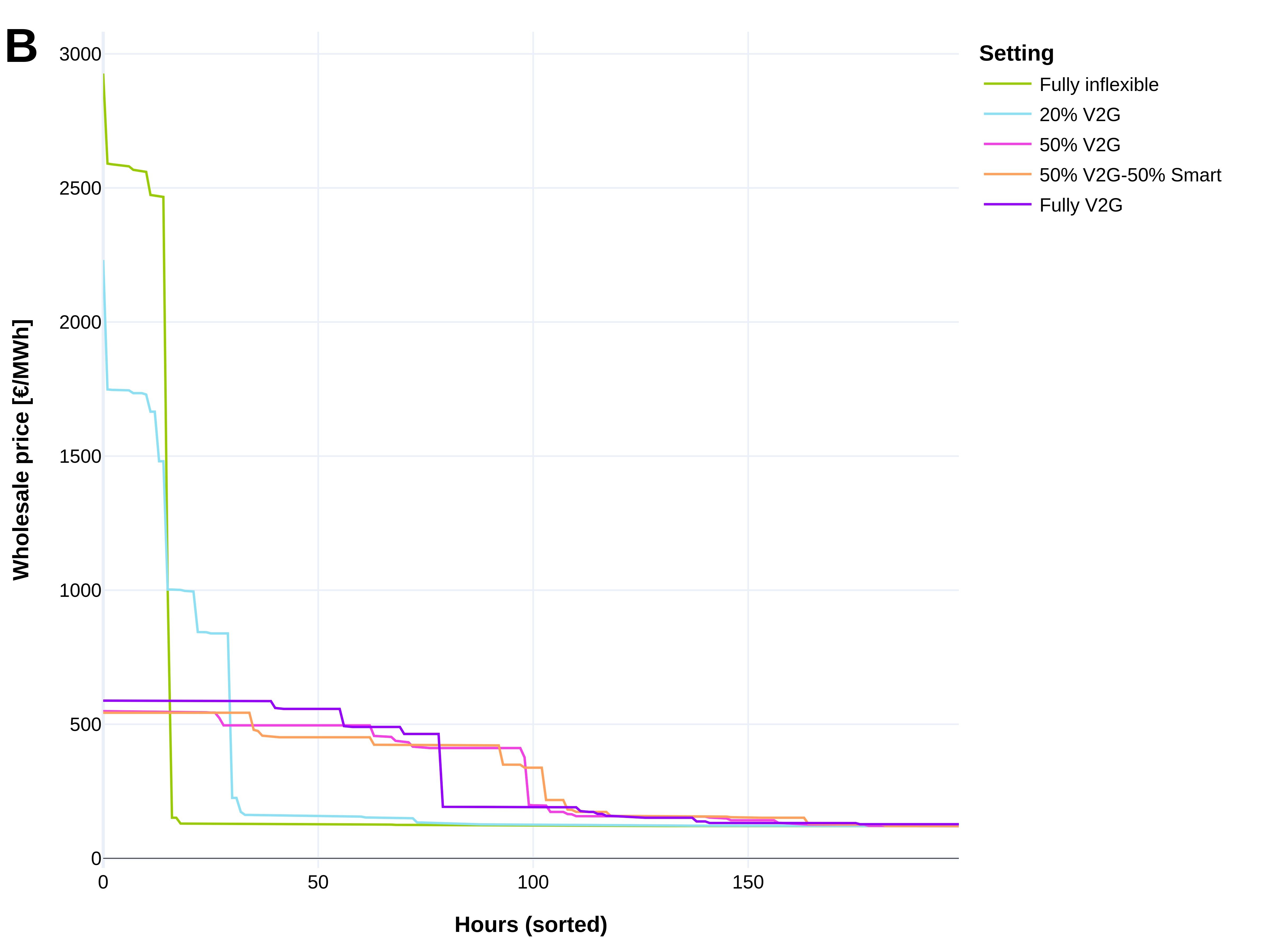}
    \includegraphics[width=0.45\linewidth]{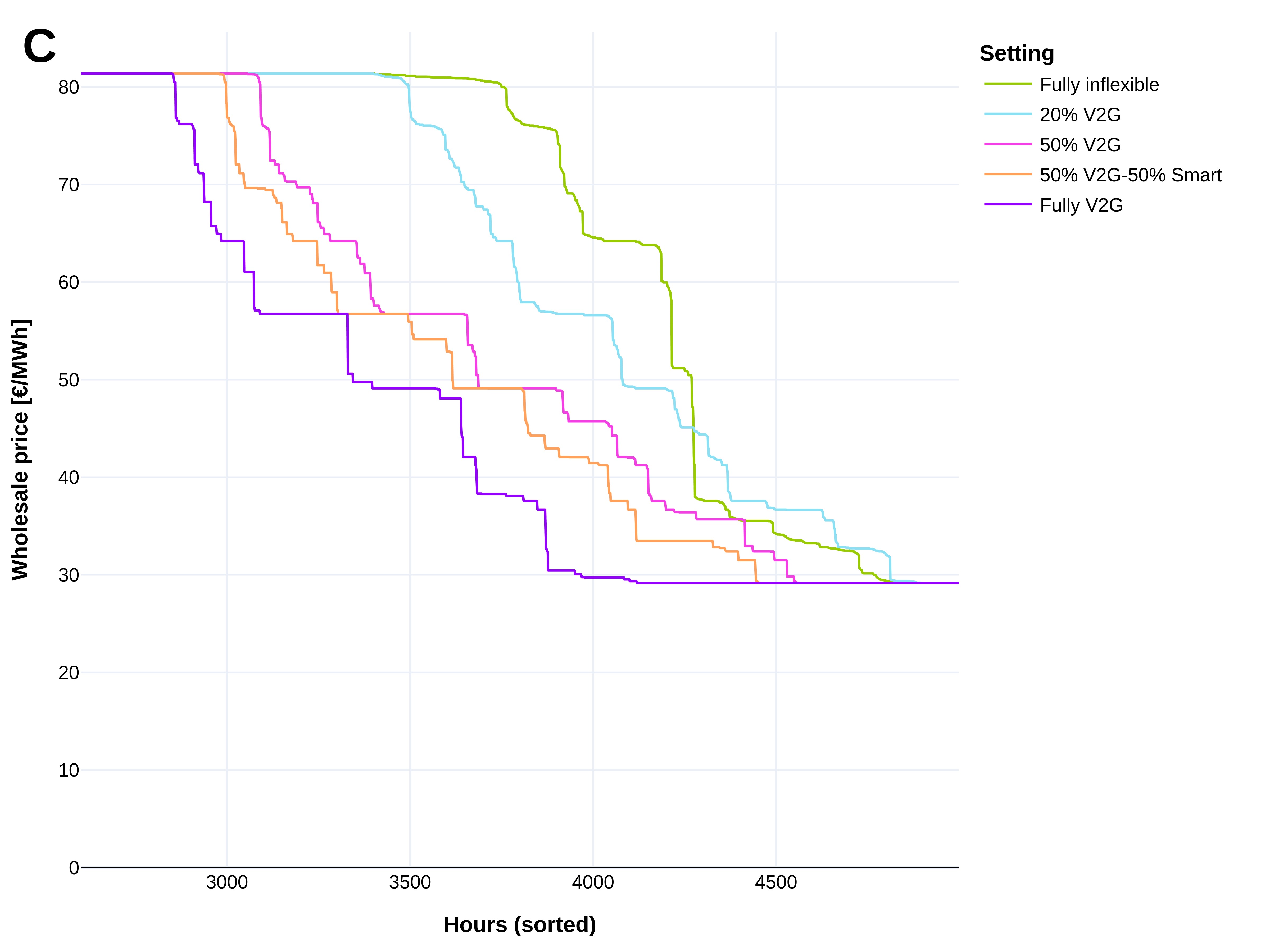}
    \caption{
    \textbf{Electricity wholesale price duration curves in Germany for a selection of scenarios and time spans.}
    Hours are sorted by decreasing price. Counterclockwise starting from the upper-left panel, selected time spans include: the full 8,760 hours (\textbf{A}); the first 200 hours (\textbf{B}); hours between 2,600 and 5,000 (\textbf{C}); the last 1,260 hours (\textbf{D}). Scenarios ``20\% V2G'' and ``50\% V2G'' refer to settings where all BEVs that are not bidirectionally charging are inflexibly charging.
    }
    \label{fig:price-duration-curves}
\end{figure}

\begin{figure}[!ht]
    \centering
    \includegraphics[width=.5\linewidth]{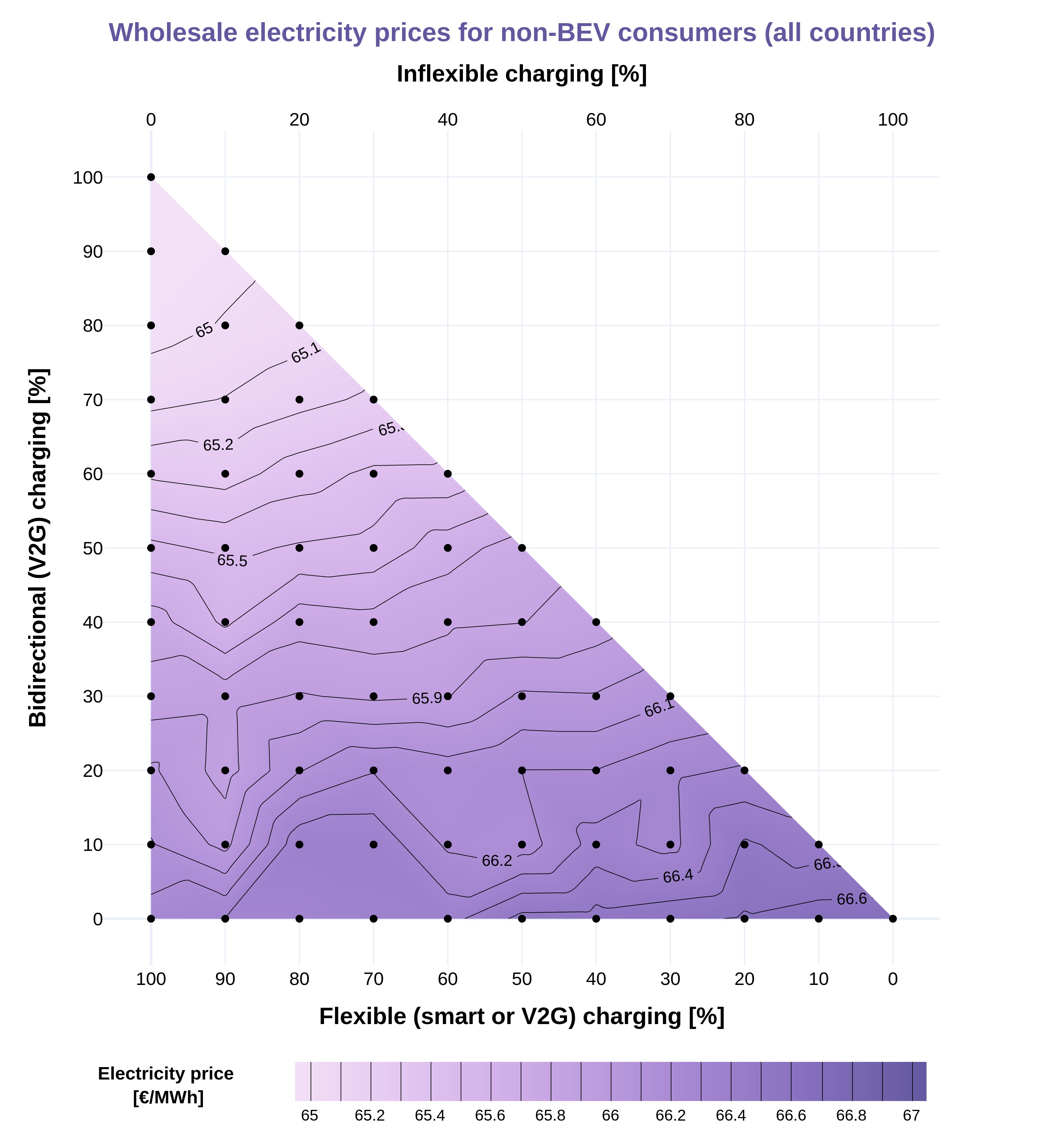}
    \caption{
    \textbf{Average wholesale electricity prices for non-BEV consumers in all countries (including Germany).} 
    Wholesale prices are weighted by the non-BEV consumer load for each hour, excluding the load for BEVs and electrolyzers.
    }
    \label{fig:heatmap_mean_baseload_price_wo_de}
\end{figure}

\clearpage

\subsection{Input data}
\begin{table}[!ht]
\centering
\resizebox{\textwidth}{!}{%
\begin{tabular}{@{}lrrrrrr@{}}
\toprule
\rowcolor[HTML]{FFFFFF} 
\multicolumn{1}{c}{\cellcolor[HTML]{FFFFFF}} &
  \multicolumn{3}{c}{\cellcolor[HTML]{FFFFFF}\textbf{Costs}} &
  \multicolumn{1}{c}{\cellcolor[HTML]{FFFFFF}} &
  \multicolumn{1}{c}{\cellcolor[HTML]{FFFFFF}} &
  \multicolumn{1}{c}{\cellcolor[HTML]{FFFFFF}} \\ \cmidrule(lr){2-4}
\rowcolor[HTML]{FFFFFF} 
\multicolumn{1}{c}{\cellcolor[HTML]{FFFFFF}} &
  \multicolumn{1}{c}{\cellcolor[HTML]{FFFFFF}Investment} &
  \multicolumn{1}{c}{\cellcolor[HTML]{FFFFFF}Fixed O\&M} &
  \multicolumn{1}{c}{\cellcolor[HTML]{FFFFFF}Variable \& fuel} &
  \multicolumn{1}{c}{\multirow{-2}{*}{\cellcolor[HTML]{FFFFFF}\textbf{Efficiency}}} &
  \multicolumn{1}{c}{\multirow{-2}{*}{\cellcolor[HTML]{FFFFFF}\textbf{Lifetime}}} &
  \multicolumn{1}{c}{\multirow{-2}{*}{\cellcolor[HTML]{FFFFFF}\textbf{Carbon content}}} \\
\rowcolor[HTML]{FFFFFF} 
\multicolumn{1}{c}{\multirow{-3}{*}{\cellcolor[HTML]{FFFFFF}\textbf{Technology}}} &
  \multicolumn{1}{c}{\cellcolor[HTML]{FFFFFF}{[}\euro/kW{]}} &
  \multicolumn{1}{c}{\cellcolor[HTML]{FFFFFF}{[}\euro/kW/a{]}} &
  \multicolumn{1}{c}{\cellcolor[HTML]{FFFFFF}{[}\euro/MWh{]}} &
  \multicolumn{1}{c}{\cellcolor[HTML]{FFFFFF}} &
  \multicolumn{1}{c}{\cellcolor[HTML]{FFFFFF}{[}years{]}} &
  \multicolumn{1}{c}{\cellcolor[HTML]{FFFFFF}{[}t/MWh{]}} \\ \midrule
\cellcolor[HTML]{FFFFFF}Lignite &
  \cellcolor[HTML]{FFFFFF}1,500.00 &
  30.00 &
  \cellcolor[HTML]{FFFFFF}13.68 &
  0.390 &
  \cellcolor[HTML]{FFFFFF}40 &
  \cellcolor[HTML]{FFFFFF}0.399 \\
\rowcolor[HTML]{FFFFFF} 
Hard coal          & 4,575.69 & 59.98  & 13.15 & 0.405 & 40 & 0.337 \\
\rowcolor[HTML]{FFFFFF} 
Other fossil fuels        & 1,500.00 & 15.00  & 15.10 & 0.350 & 25 & 0.350 \\
\rowcolor[HTML]{FFFFFF} 
Oil                & 361.55   & 8.98   & 49.54 & 0.350 & 25 & 0.257 \\
\rowcolor[HTML]{FFFFFF} 
Natural gas (CCGT) & 882.60   & 29.56  & 21.06 & 0.580 & 25 & 0.201 \\
\rowcolor[HTML]{FFFFFF} 
Natural gas (OCGT) & 467.88   & 8.24   & 26.06 & 0.410 & 25 & 0.201 \\
\rowcolor[HTML]{FFFFFF} 
Nuclear            & 8,594.14 & 109.15 & 5.24  & 0.337 & 40 & 0.000 \\
\rowcolor[HTML]{FFFFFF} 
Bioenergy          & 2,209.00 & 100.00 & 13.65 & 0.468 & 30 & 0.000 \\
\rowcolor[HTML]{FFFFFF} 
Run-of-river       & 3,000.00 & 60.00  & 0.00  & 1.000 & 50 & 0.000 \\ 
\rowcolor[HTML]{FFFFFF} 
Offshore wind      & 1,800.00 & 39.00  & 4.00  & 1.000 & 30 & 0.000 \\
\rowcolor[HTML]{FFFFFF} 
Onshore wind       & 1,146.64 & 16.66  & 1.98  & 1.000 & 30 & 0.000 \\
\rowcolor[HTML]{FFFFFF} 
Solar photovoltaic       & 380.00   & 9.50   & 0.00  & 1.000 & 40 & 0.000 \\ \bottomrule
\end{tabular}%
}
\caption{\textbf{Cost and technology assumptions for electricity generation technologies.} The assumed interest rate is 0.04 and the assumed carbon price is 130 euros per tCO$_2$eq.}
\label{tab:parameters_generation}
\end{table}
\begin{table}[!ht]
\centering
\resizebox{\textwidth}{!}{%
\begin{tabular}{@{}
>{\columncolor[HTML]{FFFFFF}}l 
>{\columncolor[HTML]{FFFFFF}}r 
>{\columncolor[HTML]{FFFFFF}}r 
>{\columncolor[HTML]{FFFFFF}}r 
>{\columncolor[HTML]{FFFFFF}}r 
>{\columncolor[HTML]{FFFFFF}}r 
>{\columncolor[HTML]{FFFFFF}}r 
>{\columncolor[HTML]{FFFFFF}}r 
>{\columncolor[HTML]{FFFFFF}}r 
>{\columncolor[HTML]{FFFFFF}}r 
>{\columncolor[HTML]{FFFFFF}}r @{}}
\toprule
\multicolumn{1}{c}{\cellcolor[HTML]{FFFFFF}} &
  \multicolumn{3}{c}{\cellcolor[HTML]{FFFFFF}\textbf{Investment costs}} &
  \multicolumn{3}{c}{\cellcolor[HTML]{FFFFFF}\textbf{Fixed costs}} &
  \multicolumn{1}{c}{\cellcolor[HTML]{FFFFFF}\textbf{Variable costs}} &
  \multicolumn{2}{c}{\cellcolor[HTML]{FFFFFF}\textbf{Efficiency}} &
  \multicolumn{1}{c}{\cellcolor[HTML]{FFFFFF}} \\ \cmidrule(lr){2-10}
\multicolumn{1}{c}{\cellcolor[HTML]{FFFFFF}} &
  \multicolumn{1}{c}{\cellcolor[HTML]{FFFFFF}Energy} &
  \multicolumn{1}{c}{\cellcolor[HTML]{FFFFFF}Charging} &
  \multicolumn{1}{c}{\cellcolor[HTML]{FFFFFF}Discharging} &
  \multicolumn{1}{c}{\cellcolor[HTML]{FFFFFF}Energy} &
  \multicolumn{1}{c}{\cellcolor[HTML]{FFFFFF}Charging} &
  \multicolumn{1}{c}{\cellcolor[HTML]{FFFFFF}Discharging} &
  \multicolumn{1}{c}{\cellcolor[HTML]{FFFFFF}} &
  \multicolumn{1}{c}{\cellcolor[HTML]{FFFFFF}Charging} &
  \multicolumn{1}{c}{\cellcolor[HTML]{FFFFFF}Discharging} &
  \multicolumn{1}{c}{\multirow{-2}{*}{\cellcolor[HTML]{FFFFFF}\textbf{Lifetime}}} \\
\multicolumn{1}{c}{\multirow{-3}{*}{\cellcolor[HTML]{FFFFFF}\textbf{Technology}}} &
  \multicolumn{1}{c}{\cellcolor[HTML]{FFFFFF}{[}EUR/kWh{]}} &
  \multicolumn{1}{c}{\cellcolor[HTML]{FFFFFF}{[}EUR/kW{]}} &
  \multicolumn{1}{c}{\cellcolor[HTML]{FFFFFF}{[}EUR/kW{]}} &
  \multicolumn{1}{c}{\cellcolor[HTML]{FFFFFF}{[}EUR/kWh/a{]}} &
  \multicolumn{1}{c}{\cellcolor[HTML]{FFFFFF}{[}EUR/kW/a{]}} &
  \multicolumn{1}{c}{\cellcolor[HTML]{FFFFFF}{[}EUR/kW/a{]}} &
  \multicolumn{1}{c}{\cellcolor[HTML]{FFFFFF}{[}EUR/MW{]}} &
  \multicolumn{1}{c}{\cellcolor[HTML]{FFFFFF}} &
  \multicolumn{1}{c}{\cellcolor[HTML]{FFFFFF}} &
  \multicolumn{1}{c}{\cellcolor[HTML]{FFFFFF}{[}years{]}} \\ \midrule
Reservoir                     & 56.03  & 0.00   & 685.81 & 0.00  & 0.00  & 30.00 & 0.10 & -     & 0.950 & 60   \\
Pumped hydro storage (closed)  & 56.03  & 685.81 & 685.81 & 13.65 & 0.00  & 0.00  & 0.50 & 0.894 & 0.894 & 60  \\
Pumped hydro storage (open)    & 56.03  & 685.81 & 685.81 & 13.65 & 0.00  & 0.00  & 0.10 & 0.894 & 0.894 & 60  \\
Stationary Li-ion battery      & 151.00 & 85.08  & 85.08  & 0.00  & 0.29  & 0.29  & 0.96 & 0.985 & 0.975 & 25  \\
Long-duration storage         &        &        &        &       &       &       &      &       &       &           \\
… PEM electrolyzer            & -      & 650.00 & -      & -     & 13.00 & -     & -    & 0.585 & -     & 25  \\
… Hydrogen cavern compressor  & -      & 80.17  & -      & -     & 3.21  & -     & 0.00 & 0.995 & -     & 15   \\
… Hydrogen cavern storage      & 2.13   & -      & -      & 0.003 & -     & -     & -    & -     & -     & 100  \\
… Hydrogen (OCGT)            & -      & -      & 538.07 & -     & -     & 8.24  & 5.00 & -     & 0.410 & 25   \\ \bottomrule
\end{tabular}%
}
\caption{\textbf{Cost and technology assumptions for electricity storage technologies.} The assumed interest rate is 0.04.}
\label{tab:parameters_storage}
\end{table}
\begin{table}[!ht]
\centering
\resizebox{\textwidth}{!}{%
\begin{tabular}{@{}
>{\columncolor[HTML]{FFFFFF}}c 
>{\columncolor[HTML]{FFFFFF}}c 
>{\columncolor[HTML]{FFFFFF}}c 
>{\columncolor[HTML]{FFFFFF}}c 
>{\columncolor[HTML]{FFFFFF}}c 
>{\columncolor[HTML]{FFFFFF}}c 
>{\columncolor[HTML]{FFFFFF}}c 
>{\columncolor[HTML]{FFFFFF}}c 
>{\columncolor[HTML]{FFFFFF}}c 
>{\columncolor[HTML]{FFFFFF}}c 
>{\columncolor[HTML]{FFFFFF}}c 
>{\columncolor[HTML]{FFFFFF}}c @{}}
\toprule
\textbf{} &
  \textbf{Austria} &
  \textbf{Belgium} &
  \textbf{Switzerland} &
  \textbf{Czech Republic} &
  \textbf{Germany} &
  \textbf{Denmark} &
  \textbf{France} &
  \textbf{Italy} &
  \textbf{Luxembourg} &
  \textbf{Netherlands} &
  \textbf{Poland} \\ \midrule
\multicolumn{1}{l|}{\cellcolor[HTML]{FFFFFF}\textbf{Austria}} &
  \textbf{-} &
  - &
  1200 &
  900 &
  7400 &
  - &
  - &
  500 &
  - &
  - &
  - \\
\multicolumn{1}{l|}{\cellcolor[HTML]{FFFFFF}\textbf{Belgium}} &
  - &
  \textbf{-} &
  - &
  - &
  1000 &
  - &
  5300 &
  - &
  180 &
  4400 &
  - \\
\multicolumn{1}{l|}{\cellcolor[HTML]{FFFFFF}\textbf{Switzerland}} &
  1200 &
  - &
  \textbf{-} &
  - &
  6600 &
  - &
  4700 &
  1700 &
  - &
  - &
  - \\
\multicolumn{1}{l|}{\cellcolor[HTML]{FFFFFF}\textbf{Czech Republic}} &
  900 &
  - &
  - &
  \textbf{-} &
  2100 &
  - &
  - &
  - &
  - &
  - &
  800 \\
\multicolumn{1}{l|}{\cellcolor[HTML]{FFFFFF}\textbf{Germany}} &
  7400 &
  1000 &
  6600 &
  2100 &
  \textbf{-} &
  44485 &
  6000 &
  - &
  2300 &
  5000 &
  3000 \\
\multicolumn{1}{l|}{\cellcolor[HTML]{FFFFFF}\textbf{Denmark}} &
  - &
  - &
  - &
  - &
  4485 &
  \textbf{-} &
  - &
  - &
  - &
  2000 &
  - \\
\multicolumn{1}{l|}{\cellcolor[HTML]{FFFFFF}\textbf{France}} &
  - &
  5300 &
  4700 &
  - &
  6000 &
  - &
  \textbf{-} &
  2180 &
  - &
  - &
  - \\
\multicolumn{1}{l|}{\cellcolor[HTML]{FFFFFF}\textbf{Italy}} &
  - &
  - &
  1700 &
  - &
  - &
  - &
  2180 &
  \textbf{-} &
  - &
  - &
  - \\
\multicolumn{1}{l|}{\cellcolor[HTML]{FFFFFF}\textbf{Luxembourg}} &
  - &
  180 &
  - &
  - &
  2300 &
  - &
  - &
  - &
  \textbf{-} &
  - &
  - \\
\multicolumn{1}{l|}{\cellcolor[HTML]{FFFFFF}\textbf{Netherlands}} &
  - &
  4400 &
  - &
  - &
  5000 &
  2000 &
  - &
  - &
  - &
  \textbf{-} &
  - \\
\multicolumn{1}{l|}{\cellcolor[HTML]{FFFFFF}\textbf{Poland}} &
  - &
  - &
  - &
  800 &
  3000 &
  - &
  - &
  - &
  - &
  - &
  \textbf{-} \\ \bottomrule
\end{tabular}%
}
\caption{\textbf{Net Transfer Capacities (NTC) in MW.}}
\label{tab:parameters_ntc}
\end{table}
\begin{landscape}
\begin{table}[!ht]
\centering
\resizebox{\textwidth}{!}{%
\begin{tabular}{@{}
>{\columncolor[HTML]{FFFFFF}}c 
>{\columncolor[HTML]{FFFFFF}}l 
>{\columncolor[HTML]{FFFFFF}}c 
>{\columncolor[HTML]{FFFFFF}}c 
>{\columncolor[HTML]{FFFFFF}}c 
>{\columncolor[HTML]{FFFFFF}}c 
>{\columncolor[HTML]{FFFFFF}}c 
>{\columncolor[HTML]{FFFFFF}}c 
>{\columncolor[HTML]{FFFFFF}}c 
>{\columncolor[HTML]{FFFFFF}}c 
>{\columncolor[HTML]{FFFFFF}}c 
>{\columncolor[HTML]{FFFFFF}}c 
>{\columncolor[HTML]{FFFFFF}}c @{}}
\toprule
\multicolumn{1}{l}{\cellcolor[HTML]{FFFFFF}} &
  \textbf{} &
  \textbf{Germany} &
  \textbf{Austria} &
  \textbf{Belgium} &
  \textbf{Switzerland} &
  \textbf{Czech Republic} &
  \textbf{Denmark} &
  \textbf{France} &
  \textbf{Italy} &
  \textbf{Luxembourg} &
  \textbf{Netherlands} &
  \textbf{Poland} \\ \midrule
\cellcolor[HTML]{FFFFFF} &
  \textit{Lignite} &
  0-9.3 &
  0.0 &
  0.0 &
  0.0 &
  0-3.9 &
  0.0 &
  0.0 &
  0.0 &
  0.0 &
  0.0 &
  0-6.3 \\
\cellcolor[HTML]{FFFFFF} &
  \textit{Hard coal} &
  0-9.8 &
  0.0 &
  0-0.6 &
  0.0 &
  0-0.4 &
  0-0.8 &
  0.0 &
  0.0 &
  0.0 &
  0.0 &
  0-9.9 \\
\cellcolor[HTML]{FFFFFF} &
  \textit{Other fossil fuels} &
  0-4.1 &
  0-0.9 &
  0-1.3 &
  0-0.9 &
  0-1.2 &
  0-0.2 &
  0-1.9 &
  0-6.0 &
  0-0.03 &
  0-3.8 &
  0-6.8 \\
\cellcolor[HTML]{FFFFFF} &
  \textit{Oil} &
  0-1.2 &
  0-0.2 &
  0.0 &
  0.0 &
  0-0.01 &
  0.0 &
  0 &
  0.0 &
  0.0 &
  0.0 &
  0.0 \\
\cellcolor[HTML]{FFFFFF} &
  \textit{Natural gas (CCGT)} &
  0-$\infty$ &
  0-2.8 &
  0-7.6 &
  0.0 &
  0-1.3 &
  0.0 &
  0-6.5 &
  0-3.9 &
  0.0 &
  0-8.6 &
  0-5.0 \\
\cellcolor[HTML]{FFFFFF} &
  \textit{Natural gas (OCGT)} &
  0-$\infty$ &
  0-0.6 &
  0-1.1 &
  0.0 &
  0.0 &
  0.0 &
  0-0.9 &
  0-5.4 &
  0.0 &
  0-0.6 &
  0.0 \\
\cellcolor[HTML]{FFFFFF} &
  \textit{Nuclear} &
  0.0 &
  0.0 &
  0.0 &
  1.2 &
  4.0 &
  0.0 &
  58.2 &
  0.0 &
  0.0 &
  0.5 &
  0.0 \\ \cmidrule(l){2-13} 
\cellcolor[HTML]{FFFFFF} &
  \textit{Bioenergy} &
  6.0 &
  0.6 &
  0.2 &
  1.2 &
  1.1 &
  0.7 &
  2.6 &
  4.9 &
  0.05 &
  0.5 &
  1.4 \\
\cellcolor[HTML]{FFFFFF} &
  \textit{Run-of-river} &
  3.9 &
  6.4 &
  0.2 &
  4.2 &
  0.4 &
  0.0 &
  13.6 &
  7.0 &
  0.04 &
  0.04 &
  0.4 \\
\cellcolor[HTML]{FFFFFF} &
  \textit{Offshore wind} &
  30.0 &
  0.0 &
  4.3 &
  0.0 &
  0.0 &
  4.8 &
  3.0 &
  0.6 &
  0.0 &
  6.7 &
  0.9 \\
\cellcolor[HTML]{FFFFFF} &
  \textit{Onshore wind} &
  0-$\infty$ &
  10.0 &
  5.9 &
  1.3 &
  3.0 &
  5.5 &
  44.1 &
  19.0 &
  0-0.3 &
  8.3 &
  11.3 \\
\multirow{-12}{*}{\cellcolor[HTML]{FFFFFF}\textbf{Generation technologies}} &
  \textit{Solar photovoltaic} &
  0-$\infty$ &
  15.0 &
  13.9 &
  11.0 &
  10.5 &
  4.7 &
  42.6 &
  49.3 &
  0.3 &
  15.5 &
  12.2 \\ \midrule
\cellcolor[HTML]{FFFFFF} &
  \textit{Reservoir} &
   &
   &
   &
   &
   &
   &
   &
   &
   &
   &
   \\
\cellcolor[HTML]{FFFFFF} &
  \multicolumn{1}{l}{\cellcolor[HTML]{FFFFFF}...power out} &
  0.8 &
  2.8 &
  0.0 &
  8.5 &
  0.2 &
  0.0 &
  9.8 &
  8.8 &
  0.0 &
  0.0 &
  0.4 \\
\cellcolor[HTML]{FFFFFF} &
  \multicolumn{1}{l}{\cellcolor[HTML]{FFFFFF}...energy {[}TWh{]}} &
  0.237 &
  0.769 &
  0.0 &
  7.912 &
  0.002 &
  0.0 &
  10.0 &
  5.568 &
  0.0 &
  0.0 &
  0.001 \\
\cellcolor[HTML]{FFFFFF} &
  \textit{Pumped hydro storage (closed)} &
   &
   &
   &
   &
   &
   &
   &
   &
   &
   &
   \\
\cellcolor[HTML]{FFFFFF} &
  \multicolumn{1}{l}{\cellcolor[HTML]{FFFFFF}...power in/out} &
  7.17/7.01 &
  0.45/0.45 &
  1.23/1.31 &
  1.90/1.90 &
  0.66/0.69 &
  0.0/0.0 &
  1.95/1.95 &
  4.17/4.17 &
  0.0/0.0 &
  0.0/0.0 &
  1.49/1.32 \\
\cellcolor[HTML]{FFFFFF} &
  \multicolumn{1}{l}{\cellcolor[HTML]{FFFFFF}...energy {[}TWh{]}} &
  0.392 &
  0.004 &
  0.006 &
  0.056 &
  0.004 &
  0.0 &
  0.010 &
  0.061 &
  0.0 &
  0.0 &
  0.006 \\
\cellcolor[HTML]{FFFFFF} &
  \textit{Pumped hydro storage (open)} &
   &
   &
   &
   &
   &
   &
   &
   &
   &
   &
   \\
\cellcolor[HTML]{FFFFFF} &
  \multicolumn{1}{l}{\cellcolor[HTML]{FFFFFF}...power in/out} &
  1.86/2.14 &
  5.33/5.61 &
  0.0/0.0 &
  1.89/2.46 &
  0.60/0.65 &
  0.0/0.0 &
  1.85/1.85 &
  2.22/3.62 &
  0.0/0.0 &
  0.0/0.0 &
  0.17/0.22 \\
\cellcolor[HTML]{FFFFFF} &
  \multicolumn{1}{l}{\cellcolor[HTML]{FFFFFF}...energy {[}TWh{]}} &
  0.471 &
  1.747 &
  0.0 &
  1.194 &
  0.003 &
  0.0 &
  0.090 &
  0.290 &
  0.0 &
  0.0 &
  0.001 \\
\cellcolor[HTML]{FFFFFF} &
  \textit{Stationary Li-ion battery} &
   &
   &
   &
   &
   &
   &
   &
   &
   &
   &
   \\
\cellcolor[HTML]{FFFFFF} &
  \multicolumn{1}{l}{\cellcolor[HTML]{FFFFFF}...power in/out} &
  0-$\infty$/0-$\infty$ &
  0-$\infty$/0-$\infty$ &
  0-$\infty$/0-$\infty$ &
  0-$\infty$/0-$\infty$ &
  0-$\infty$/0-$\infty$ &
  0-$\infty$/0-$\infty$ &
  0-$\infty$/0-$\infty$ &
  0-$\infty$/0-$\infty$ &
  0-$\infty$/0-$\infty$ &
  0-$\infty$/0-$\infty$ &
  0-$\infty$/0-$\infty$ \\
\cellcolor[HTML]{FFFFFF} &
  \multicolumn{1}{l}{\cellcolor[HTML]{FFFFFF}...energy {[}TWh{]}} &
  0-$\infty$ &
  0-$\infty$ &
  0-$\infty$ &
  0-$\infty$ &
  0-$\infty$ &
  0-$\infty$ &
  0-$\infty$ &
  0-$\infty$ &
  0-$\infty$ &
  0-$\infty$ &
  0-$\infty$ \\ \cmidrule(l){2-13} 
\cellcolor[HTML]{FFFFFF} &
  \textit{PEM electrolyzer} &
   &
   &
   &
   &
   &
   &
   &
   &
   &
   &
   \\
\cellcolor[HTML]{FFFFFF} &
  \multicolumn{1}{l}{\cellcolor[HTML]{FFFFFF}...power} &
  0-$\infty$ &
  0-$\infty$ &
  0-$\infty$ &
  0-$\infty$ &
  0-$\infty$ &
  0-$\infty$ &
  0-$\infty$ &
  0-$\infty$ &
  0-$\infty$ &
  0-$\infty$ &
  0-$\infty$ \\
\cellcolor[HTML]{FFFFFF} &
  \textit{Hydrogen cavern} &
   &
   &
   &
   &
   &
   &
   &
   &
   &
   &
   \\
\cellcolor[HTML]{FFFFFF} &
  \multicolumn{1}{l}{\cellcolor[HTML]{FFFFFF}...energy {[}TWh{]}} &
  35731.0 &
  0.0 &
  0.0 &
  0.0 &
  0.0 &
  7698.3 &
  511.0 &
  0.0 &
  0.0 &
  10420.9 &
  7256.9 \\
\cellcolor[HTML]{FFFFFF} &
  \textit{Hydrogen (OCGT)} &
   &
   &
   &
   &
   &
   &
   &
   &
   &
   &
   \\
\multirow{-18}{*}{\cellcolor[HTML]{FFFFFF}\textbf{Storage technologies}} &
  \multicolumn{1}{l}{\cellcolor[HTML]{FFFFFF}...power} &
  0-$\infty$ &
  0 &
  0 &
  0 &
  0 &
  0-$\infty$ &
  0-$\infty$ &
  0 &
  0 &
  0-$\infty$ &
  0-$\infty$ \\ \bottomrule
\end{tabular}%
}
\caption{\textbf{Capacity bounds by country and technology.} Based on ENTSO-E. ``TYNDP 2018. Project Sheets''. Tech. rep. 2018. Unless otherwise specified, capacity bounds are expressed in gigawatts (GW). For a given technology and country, two numbers separated by an hyphen refer to endogenous capacity investments within this range.}
\label{tab:parameters_capacitybounds}
\end{table}
\end{landscape}
\begin{table}[!ht]
\centering
\begin{tabular}{@{}
>{\columncolor[HTML]{FFFFFF}}r 
>{\columncolor[HTML]{FFFFFF}}c @{}}
\toprule
\multicolumn{1}{c}{\cellcolor[HTML]{FFFFFF}\textbf{Country}} & \textbf{Electricity demand {[}TWh{]}} \\ \midrule
\textit{Luxembourg}     & 8.8   \\
\textit{Denmark}        & 52.6  \\
\textit{Switzerland}    & 64.0  \\
\textit{Czech Republic} & 74.1  \\
\textit{Austria}        & 82.2  \\
\textit{Belgium}        & 95.2  \\
\textit{Netherlands}    & 139.9 \\
\textit{Poland}         & 180.9 \\
\textit{Italy}          & 330.6 \\
\textit{France}         & 478.0 \\
\textit{Germany}        & 583.1 \\ \bottomrule
\end{tabular}
\caption{\textbf{Yearly electricity demand of non-BEV consumers by country in TWh.} Data are taken from the ENTSO-E Pan-European Climate Database (PECD 2021.3).}
\label{tab:parameters_load}
\end{table}
\end{document}